\documentclass[superscriptaddress,amsmath,amssymb, aps, prx,longbibliography,twocolumn, footinbib]{revtex4-2}
\usepackage{upgreek}
\usepackage{color}
\usepackage{graphicx}
\usepackage{nccmath}
\usepackage{bm}
\usepackage{hyperref}
\hypersetup{colorlinks,breaklinks,
            urlcolor=[rgb]{0,0,0.36},
            linkcolor=[rgb]{0,0,0.36},
            citecolor=[rgb]{0,0,0.36}}
\usepackage[mathlines]{lineno}
\usepackage{dcolumn}
\usepackage{booktabs} 
\newcommand{\ra}[1]{\renewcommand{\arraystretch}{#1}}
\usepackage{array}
\usepackage{soul}

\newcommand{\Harvard}{Department of Physics, Harvard University, Cambridge, Massachusetts 02138, USA.}

\newcommand{\HarvardAppl}{John A. Paulson School of Engineering and Applied Sciences, Harvard University, Cambridge, MA 02138.}

\newcommand{\Cal}{University of California at Berkeley, Department of Chemistry, Berkeley, California 94720, USA.}

\newcommand{\MaxP}{Max Planck Institute for the Structure and Dynamics of Matter, Luruper Chaussee 149, 22761 Hamburg, Germany.}

\newcommand{\Oxford}{Clarendon Laboratory, University of Oxford, Parks Road, Oxford OX1 3PU, UK.}

\newcommand{\Technion}{Physics Department, Technion, 32000 Haifa, Israel.}

\newcommand{\ETH}{Institute for Theoretical Physics, ETH Zurich, 8093 Zurich, Switzerland.}
\begin{document}
%\linenumbers

% ---- Title updated on 1/1/2021
% \title{Coherent dynamics due to light-superconductor interaction after incoherent photoexcitation}
% \title{Oscillatory dynamics in superconductors after incoherent photoexcitation}
\title{Periodic dynamics in superconductors induced by an impulsive optical quench}

\author{Pavel~E.~Dolgirev}\email[Correspondence to: ]{p\_dolgirev@g.harvard.edu}
\affiliation{\Harvard}
\author{Alfred~Zong}
\affiliation{\Cal}
\author{Marios~H.~Michael}
\affiliation{\Harvard}
\author{Jonathan~B.~Curtis}
\affiliation{\Harvard}
\affiliation{\HarvardAppl}
\author{Daniel Podolsky}
\affiliation{\Technion}
\author{Andrea Cavalleri}
\affiliation{\MaxP}
\affiliation{\Oxford}
\author{Eugene~Demler}
%\affiliation{\Harvard}
\affiliation{\ETH}

\date{\today}

\begin{abstract}
A number of experiments have evidenced signatures of enhanced superconducting correlations after photoexcitation. Initially, these experiments were interpreted as resulting from quasi-static changes in the Hamiltonian parameters, for example, due to lattice deformations or melting of competing phases. Yet, several recent observations indicate that these conjectures are either incorrect or do not capture all the observed phenomena, which include reflectivity exceeding unity, large shifts of Josephson plasmon edges, and appearance of new peaks in terahertz reflectivity. These observations can be explained from the perspective of a Floquet theory involving a periodic drive of system parameters, but the origin of the underlying oscillations remains unclear. In this paper, we demonstrate that following incoherent photoexcitation, long-lived oscillations are generally expected in superconductors with low-energy Josephson plasmons, such as in cuprates or fullerene superconductor K$_3$C$_{60}$. These oscillations arise from the parametric generation of plasmon pairs due to pump-induced perturbation of the superconducting order parameter. We show that this bi-plasmon response can persist even above the transition temperature as long as strong superconducting fluctuations are present. Our analysis offers a robust framework to understand light-induced superconducting behavior, and  the predicted bi-plasmon oscillations can be directly detected using available experimental techniques. 
\end{abstract}

\maketitle

\section{Introduction}

Optical manipulation of materials has emerged as a powerful tool in engineering material properties on demand. A striking instance is the phenomenon of photo-induced superconductivity, in which superconductor-like behavior is observed after photoexcitation at temperatures far higher than the equilibrium transition temperature~\cite{Cavalleri.2018}. This effect has now been demonstrated experimentally in many different systems, including high-$T_{c}$ cuprates~\cite{Hu14,Kaiser14,Fausti.2011,Nicoletti.2014,cremin2019photoenhanced,Rajasekeran.2018,Zhang.2018}, 
%YBa$_2$Cu$_3$O$_{6+x}$~\cite{Hu14,Kaiser14,Fausti.2011} and La$_{2-x}$Ba$_x$CuO$_4$ \cite{Nicoletti.2014}, 
iron-based superconductors~\cite{Suzuki.2019}, fullerene superconductor $\rm K_3 C_{60}$ \cite{mitrano2016possible,Budden21}, and organic superconductor (BEDT-TTF)$_2$Cu[N(CN)$_2$]Br \cite{buzzi20}. However, a theoretical explanation for this effect in each system is still subject to intense debate, with most interpretations coming essentially in three flavors. 

The first class of theoretical ideas suggests that photoexcitation of the material leads to quasi-static modifications of the effective Hamiltonian~\cite{mankowsky2014nonlinear,ido2017correlation,Kim.2016,Sentef.2016a,Schutt.2018,PhysRevLett.123.030603,buvca2019non,Sun.2020,Tindall.2020}: examples include light distortion of the crystal lattice likely favoring superconductivity~\cite{mankowsky2014nonlinear} or melting of a competing charge density wave (CDW)~\cite{cremin2019photoenhanced,Deghani.2020,Patel.2016,Sentef.2017}, which can even lead to metastable superconductivity~\cite{cremin2019photoenhanced}. The second approach argues that optical pumping provides a cooling mechanism for quasiparticles that allows signatures of superconductivity to persist up to higher temperatures~\cite{Robertson.2009,Robertson.2011,Tikhonov.2018,Nava.2018,Li.2020,Hoppner.2015,Denny.2015}. The third interpretation has a Floquet-like out-of-equilibrium character~\cite{Komnik.2016,Murakami.2016,Kennes.2017,Knap2016,Babadi2017,marios2020,Buzzi21,dai2021photo,Peronaci.2020,Lemonik.2018,Lemonik.2019,Okamoto2016,marios2020,Sentef.2017b,Gao.2020,Raines.2015,Okamoto2017,Schlawin.2017}, where photoexcitation results in a parametric amplification of superconducting fluctuations.

Regarding the first category of theories, several experimental observations suggest that  photo-induced superconductivity involves phenomena beyond simple modifications of the effective static Hamiltonian, and dynamical aspects play a crucial role. Firstly, in the light-induced superconducting state of K$_3$C$_{60}$, the reflection coefficient exceeded unity. This observation indicates a type of light amplification not accessible in equilibrium~\cite{Buzzi21}. Secondly, in the superconducting state of YBCO, pump-induced changes in reflectivity included a new peak at a frequency higher than the equilibrium Josephson plasmon (JP) edge \cite{Hu14}. Thirdly, in the pseudogap state of YBCO, pumping leads to the appearance of the JP edge-like feature even though in equilibrium reflection coefficient appeared featureless~\cite{mankowsky2017optically}. Finally, recent experiments %using the second harmonic generation 
demonstrated exponential growth of JPs in YBCO following pump pulse both below and above $T_c$ \cite{vanHoegen19}.

Although the second class of ideas embodies a non-equilibrium character, to date, these ideas fail to provide a simple, intuitive interpretation of the mentioned experiments. On the other hand, all the experimental observations are well understood from the third theoretical approach that is based on periodically driven systems. Specifically, one starts by assuming that the pump pulse excites a collective mode, which in turn acts as a parametric drive to amplify low-energy plasmon excitations. From the perspective of nonlinear non-equilibrium optics, using the Floquet generalization of the Fresnel formalism, one can readily explain light amplification, enhancement of overdamped JPs, and appearance of new peaks and Fano-like features in reflectivity \cite{marios2020, Buzzi21,michael2021generalized}.

The key to this Floquet picture is the existence of a coherently oscillating collective mode. Possible candidates can be a phonon that has been resonantly excited by light or an excited Higgs mode~\cite{Buzzi21,Krull.2016,Juraschek.2020}, representing an order parameter amplitude fluctuation. In $d$-wave cuprate superconductors, the Higgs mode is unlikely an option because it is expected to have high energy and to be strongly damped due to nodal quasiparticles (for dynamical aspects related to the Higgs mode, see Refs.~\cite{Yang.2020,Barlas.2013,Krull.2016,Muller.2019,Nosarzewski.2017,Manske.2020,Katsumi.2020,Juraschek.2020}). %However, in $d$-wave cuprate superconductors where light-induced superconductivity was first reported, the Higgs mode is unlikely an option because it is expected to have high energy and to be strongly damped due to nodal quasiparticles (for dynamical aspects related to the Higgs mode, see Refs.~\cite{Yang.2020,Barlas.2013,Krull.2016,Muller.2019,Nosarzewski.2017,Manske.2020,Katsumi.2020,Juraschek.2020}). 
Therefore, the crucial question is whether, without resonantly exciting relevant collective modes, we can still expect well-defined low-frequency oscillations that follow impulsive photoexcitation, such as commonly used optical pumping at 1.55~eV~\cite{Nicoletti.2014,cremin2019photoenhanced,Suzuki.2019,Zhang.2018}. The central result of our paper is that in systems with low-energy Josephson plasmons, such as in cuprates where the $c$-axis plasmon gap is in the terahertz range, long-lived oscillations generally occur even without resonant driving. Remarkably, the conversion of an incoherent pump to periodic dynamics is found both below and above the superconducting transition temperature, in both isotropic and anisotropic superconductors.

\begin{figure}[t!]
	\centering
	\includegraphics[width=0.97\linewidth]{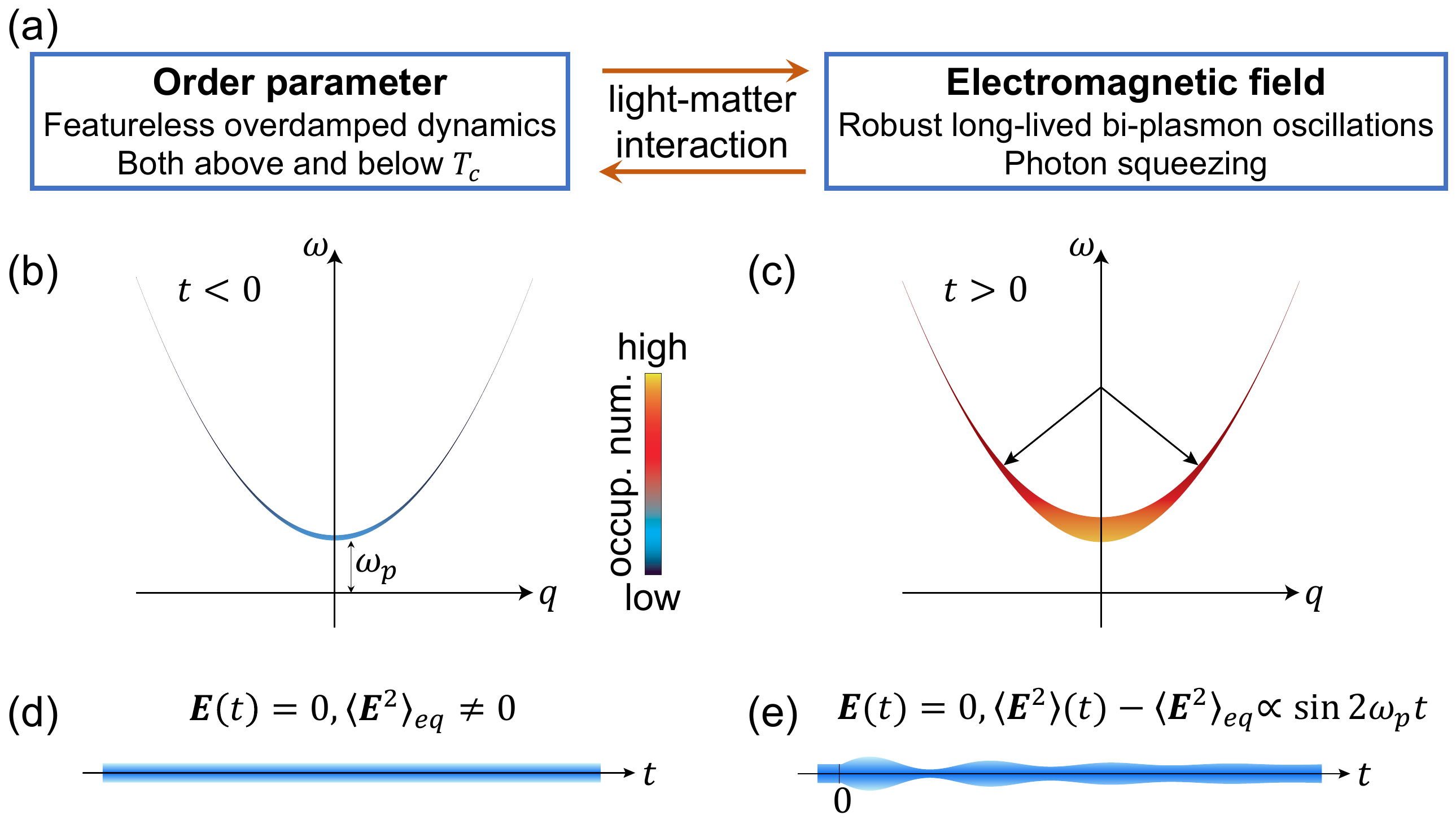}
	\caption{ (a)  The two building blocks of the effective description of non-equilibrium superconductors are the order parameter and  the electromagnetic field in the sample, and the two entities are coupled. Upon photoexcitation, the electromagnetic field displays periodic fluctuation evolution, even if the order parameter dynamics is overdamped. (b,c)~Schematic spectra of plasmons, which are gapped due to the Anderson-Higgs mechanism. Prior to photoexcitation (b), the plasmon distribution function is thermal; the thin blue curve captures the low occupation numbers. After photoexcitation (c), these plasmons proliferate (shown by the thick red-orange curve) through the creation of momentum-conserving pairs. (d,e)~Schematics of time evolution of the electric field in the sample, which always has a zero expectation value. In equilibrium (d), the electric field has a nonzero thermal variance $\langle\bm E^2\rangle_{eq}$. After an impulsive optical quench (e), this variance becomes periodically modulated, with the frequency $2\omega_p$ being twice the plasmon gap.
	}
\label{fig:Main}
\end{figure}

Before we turn to a detailed analysis, we first provide a physical picture of the mechanism that yields these oscillations despite incoherent photoexcitation. The key ingredient is plasmons in superconductors, representing fluctuations of the order parameter phase coupled to fluctuations of the electromagnetic field in the sample. This coupling renders plasmons to acquire a gap (Fig.~\ref{fig:Main}) through the Anderson-Higgs mechanism~\cite{PhysRev.112.1900}, which is closely related to the Meissner effect in static systems. In conventional superconductors, the plasmon gap is a large energy scale, it exceeds the quasiparticle gap, so that one often disregards the phase fluctuations. However, this scenario does not hold in a number of superconductors that exhibit light-induced superconducting behavior. For example, in $\rm K_3 C_{60}$, an isotropic three-dimensional superconductor, the plasmon gap is only about $20\,$THz~\cite{Buzzi21}. Thus, the phase fluctuations might be essential there~\cite{Ren.2020,jotzu2021superconducting}. Besides, in cuprates, the strong anisotropy due to the layered crystal structure dramatically renormalizes the $c$-axis plasmon gap to be of the order of $1-2\,$THz, rendering Josephson plasmons the primary lowest-energy excitations.

Motivated by $d$-wave cuprate superconductors, here we restrict the discussion to overdamped order parameter dynamics so that the Higgs-amplitude mode is strongly damped in our modeling. Despite this, we will demonstrate that the electromagnetic \emph{fluctuations} in the sample exhibit coherent, oscillatory dynamics as they couple to the order parameter phase [Fig.~\ref{fig:Main}(a)]. We consider photoexcitation with an ultrashort laser pulse, which is not tuned to a specific frequency and is not resonant with any eigenmode in the original system. The laser frequency is assumed to be much larger than any frequency of the relevant collective modes. Such an impulsive optical quench couples only to symmetry-even modes and, in particular, results in the proliferation of plasma fluctuations through the generation of momentum conserving plasmon pairs [Fig.~\ref{fig:Main}(c)]. Remarkably, even though the average electric field in the sample remains zero, the electromagnetic energy density exhibits well-defined oscillations at a frequency twice the plasmon gap [Fig.~\ref{fig:Main}(e)]. Such oscillations are well understood as two-plasmon impulsive stimulated Raman scattering, similar to what has been done for pairs of squeezed phonons in ferroelectrics~\cite{garrett1997vacuum}. The plasmons do not merely ``heat up'' after the photoexcitation, as their vacuum state is also being coherently squeezed [Fig.~\ref{fig:Main}(d) and~(e)]. Physically, this phenomenon is analogous to exciting electromagnetic modes in a cavity with moving mirrors \cite{moore1970quantum,lahteenmaki2013dynamical,Marios19}. In the latter case, the system builds up strong electric field fluctuations with random phases. As a result, averaging over ensemble gives zero expectation value of the field but a nonzero energy density. A special feature of our system is the finite gap in the plasmon spectrum. Even though a whole continuum of $(k,-k)$-plasmon pairs gets excited, the total energy density picks up the net periodic dynamics with frequency twice the spectrum gap.

To study the post-pulse evolution in both isotropic and anisotropic superconductors, we employ the time-dependent Ginzburg-Landau (TDGL) equation for overdamped stochastic dynamics of the superconducting order parameter \cite{Hohenberg1977,kopnin2001theory}. We restore the gauge invariance through minimal coupling and use Maxwell's equations for the electromagnetic field dynamics. We derive self-consistent dynamical equations for the order parameter, electromagnetic field, and their fluctuations to solve the stochastic equations. In this work, we aim to provide a  physical picture of the interplay of light and matter degrees of freedom rather than to present a full microscopic description of the post-pulse evolution, which goes beyond our effective theory and requires the knowledge of, for instance, the dynamics of the quasiparticle distribution function~\cite{kopnin2001theory}.

The TDGL approach has been widely used to understand non-equilibrium properties of superconductors~\cite{larkin2005theory,PhysRevLett.111.057002,PhysRevLett.120.117001,PhysRevB.26.4883,PhysRevLett.93.160401,PhysRevLett.96.097005,amin2004wigner,PhysRevLett.94.170402,PhysRevB.101.054203,PhysRevLett.123.097601,PhysRevLett.103.075301,PhysRevA.91.033628,PhysRevB.100.104521,PhysRevLett.127.227401}. In contrast to previous theoretical works, which neglect electromagnetic fluctuations because of a large plasmon gap, we treat the order parameter and the electromagnetic field on an equal footing because important classes of superconductors that exhibit light-induced superconductivity above $T_c$ have a relatively small plasmon gap. This model is a non-perturbative theory that takes into account fluctuations at all relevant length scales and allows us to make experimentally verifiable predictions for a broad class of superconductors. In the next section, we outline in detail our theoretical approach.

\section{Theoretical Formalism}

\subsection{Equations of motion}

To describe light-driven dynamics in layered materials, we employ a two-fluid model of superconductivity, where the superconducting fluid is coupled to the normal fluid of electrons through Maxwell's equations~\cite{Tinkham}.

\emph{Superconducting fluid---}The spontaneous symmetry breaking is described via the anisotropic gauge-invariant Ginzburg-Landau free energy (throughout the paper, we set $\hbar = k_B = 1$):
\begin{align}
     {\cal F}[\psi] & =  \int d^3 {\bm r} \Big[ \frac{1}{2m_{ab}} \Big| \Big(-i\nabla_{ab}  - \frac{e^*}{c} {\bm A}_{ab} \Big) \psi \Big|^2  \label{eqn:F}\\
     &+ \frac{1}{2m_{c}} \Big| \Big(-i\partial_z  - \frac{e^*}{c} A_z \Big) \psi \Big|^2  + \alpha |\psi|^2 + \frac{\beta}{2} |\psi|^4
    \Big].\notag
\end{align}
Here $\alpha$ and $\beta$ are the usual Landau parameters of the Mexican hat potential. $m_{ab}$ and $m_c$ are the in-plane and out-of-plane Cooper pair masses, respectively. $e^* = 2 e$ is the Cooper pair charge. The anisotropy parameter is defined as $\gamma = \sqrt{m_c/m_{ab}}$: $\gamma \gg 1$ in cuprates and $\gamma = 1$ for isotropic superconductors such as K$_3$C$_{60}$. In Eq.~\eqref{eqn:F}, the gauge-invariance is imposed through minimal coupling to the vector potential $\bm A = (\bm A_{ab}, A_z)$. The order parameter dynamics is assumed to be overdamped (model A in the classification of Ref.~\cite{Hohenberg1977}):
\begin{align}
    \tau \Big[\partial_t + i \Big(e^* \phi +  \delta \mu \Big) & \Big]  \psi({\bm r},t) = -  \frac{\delta {\cal F}}{\delta \psi^*({\bm r},t)} 
    + \eta({\bm r},t),\label{eqn:TDGL}
\end{align}
where the noise correlation function reads:
\begin{align}
    \langle \eta^*({\bm r}_1,t_1) \eta({\bm r}_2,t_2) \rangle = 2 T \tau \delta( {\bm r_1 - \bm r_2})\delta(t_1-t_2).\label{eqn:eta_noise}
\end{align}
Here $\tau$ is a dimensionless coefficient characterizing the order parameter relaxation time. $\phi$ is the scalar potential; we choose the gauge where it is zero. The term with $\delta \mu = \chi^{-1} \delta \rho$, called electrochemical potential, describes the coupling between charge fluctuations $\delta \rho$ and the order parameter phase. $\chi$ represents the compressibility, a phenomenological parameter in our approach. The superconducting electric current density is given by:
\begin{gather}
    {j}_{s, \alpha} = \hat{m}^{-1}_{\alpha\beta}  \Big[\frac{e^*}{2} \psi^* \Big(-i\partial_\beta - \frac{e^*}{c} { A}_{\beta}\Big) \psi + c.c.\Big] \label{eqn:j_s},
\end{gather}
where $\hat{m}^{-1} = \text{diag}(m_{ab}^{-1},m_{ab}^{-1},m_{c}^{-1})$ defines the mass tensor.

\emph{Normal fluid---}The normal current density reads
\begin{gather}
    {\bm j}_n = \hat{\sigma} ( {\bm E} - \nabla \delta \mu/e^*) + \bm \xi,
\end{gather}
where $\hat{\sigma} = \text{diag} (\sigma_{ab},\sigma_{ab},\sigma_{c})$ is the normal-state conductivity tensor, assumed to be momentum- and frequency-independent. We included the Johnson-Nyquist noise to enforce the fluctuation-dissipation theorem~\cite{kamenev2011field}:
\begin{gather}
\langle\xi_\alpha({\bm r}_1,t_1) \xi_\beta({\bm r}_2,t_2) \rangle  = 2 \sigma_{\alpha \beta}  T \delta( {\bm r_1 - \bm r_2})\delta(t_1-t_2).\label{eqn:JN_noise}
\end{gather}
Note that we neglected the relaxation of the normal current density, an approximation valid at low frequencies $\omega \lesssim \tau_{\rm mf}^{-1}$, where $\tau_{\rm mf}$ is the mean-free time. The charge conservation is expressed through the continuity equation:
\begin{gather}
    \frac{\partial \delta\rho}{\partial t} + \nabla\cdot ({\bm j}_n + {\bm j}_s) = 0.\label{eqn:cnt}
\end{gather}

\emph{The Maxwell equations---}In the gauge where the scalar potential is zero: ${\bm E} = - \frac{1}{c} \partial_t {\bm A}$ and ${\bm B} = \nabla \times {\bm A}$. The remaining Maxwell equations read:
\begin{gather}
 \nabla \cdot {\bm E} = 4\pi \delta\rho, \label{eqn:ME_2}\\
    \nabla \times {\bm B} = \frac{1}{c}  \frac{\partial {\bm E}}{\partial t} +  \frac{4\pi}{c}({\bm j}_n + {\bm j}_s). \label{eqn:ME_4}
\end{gather}

\emph{Theory of dynamical Gaussian fluctuations---}The above equations of motion can be formulated as stochastic first-order equations on eight real physical degrees of freedom $\bm q = (\psi_1,\psi_2,\bm A, \bm E)$, where $\psi = \psi_1 + i \psi_2$. One can then write a single first-order Fokker-Planck equation on the cumulative distribution functional ${\cal P}[t; \bm q]$. One of the simplifications we employ in this work is that we assume that ${\cal P}$ remains Gaussian throughout the time evolution, i.e. ${\cal P}$ is fully characterized by time-dependent expectation values $\langle \bm q\rangle(t)$ and {\it instantaneous} correlators $\langle q_\alpha(\bm r_1, t) q_\beta(\bm r_2, t) \rangle_c$. We further assume that the system is homogeneous in space so that in momentum space, one only has correlators of type $\langle q_\alpha(-\bm k, t) q_\beta(\bm k, t) \rangle_c$. The main technical result of this work is that we derive the first-order equations of motion on all of those correlators. The derivation, together with the final form of equations of motion, are relegated to Appendix~\ref{Sec:Main_eqn}. The obtained set of equations allows us to investigate fluctuating dynamics after photoexcitation events numerically.

\subsection{Equilibrium collective modes}

To understand how photoexcitation leads to coherent dynamics involving pairs of plasmons, we first establish the frequency of plasmons. %, which may even exist above $T_c$ due to superconducting fluctuations. 
Simple mean-field analysis shows that in anisotropic superconductors, the equilibrium $c$-axis JP gap is given by (see Appendix~\ref{sec: Collective modes}):
\begin{align}
    \omega_c =  \sqrt{ \frac{4\pi (e^*)^2}{\gamma^2 m_{ab}}  |\langle\psi\rangle|^2 - (2\pi \sigma_c)^2} - 2\pi \sigma_c i. \label{eqn_plasmon}
\end{align}
Here $\langle \psi\rangle $ is the order parameter expectation value. This formula suggests that plasmons are present only below $T_c$, where $\langle \psi\rangle \neq 0$. However, one does expect that JPs constitute collective modes of the system above $T_c$ as well. Within the Gaussian approximation, the plasmon frequency is given by a formula similar to Eq.~\eqref{eqn_plasmon} but with $|\langle\psi\rangle|^2$ replaced by $\langle|\psi|^2\rangle$. This equilibrium average of the square of the order parameter amplitude contains contributions from both the long-range expectation value $\langle \psi\rangle$ as well as from superconducting fluctuations. The Gaussian approximation, however, underestimates the JP dephasing arising from statistical fluctuations of the phase of the local superconducting order parameter. In the context of JPs in the vortex liquid state, this question has been analyzed by Koshelev and Bulaevskii~\cite{PhysRevB.60.R3743}, who found that the lifetime of plasmons becomes shorter as the superconducting correlation length decreases.

Following these analyses, we expect to find three regimes of JP dynamics, as shown in Fig.~\ref{fig:Quenches}(a). i) For $T\leq T_c$, the plasmons develop due to the non-zero order parameter expectation value $\langle \psi\rangle \neq 0$. ii) For moderate temperatures above $T_c$, $T_c < T \leq T^*$, the long-range coherence is absent $\langle \psi\rangle =0$, but strong superconducting fluctuations, as measured by $\langle |\psi|^2\rangle$, provide large local superconducting amplitude and sufficiently long correlation length. In this case, we expect to find JPs with a frequency that is still larger than their decay rate. In the discussion below, we employ the Gaussian approximation to analyze plasmon dynamics in this regime, which we call pseudogap. Plasmon dissipation arising from scattering on statistical fluctuations of the order parameter~\cite{PhysRevB.60.R3743} can be included by renormalizing the effective normal fluid conductivity $\sigma_c$. iii) For $T>T^*$, plasmons become strongly overdamped. 
We remark that frequencies of the two other $ab$-plasmons are determined by the in-plane superfluid density, and for $\gamma \gg 1$, they should have much larger energies and hence stronger damping. For this reason, we will not discuss them in the current paper.

\begin{figure}[t!]
	\centering
	\includegraphics[width=1\linewidth]{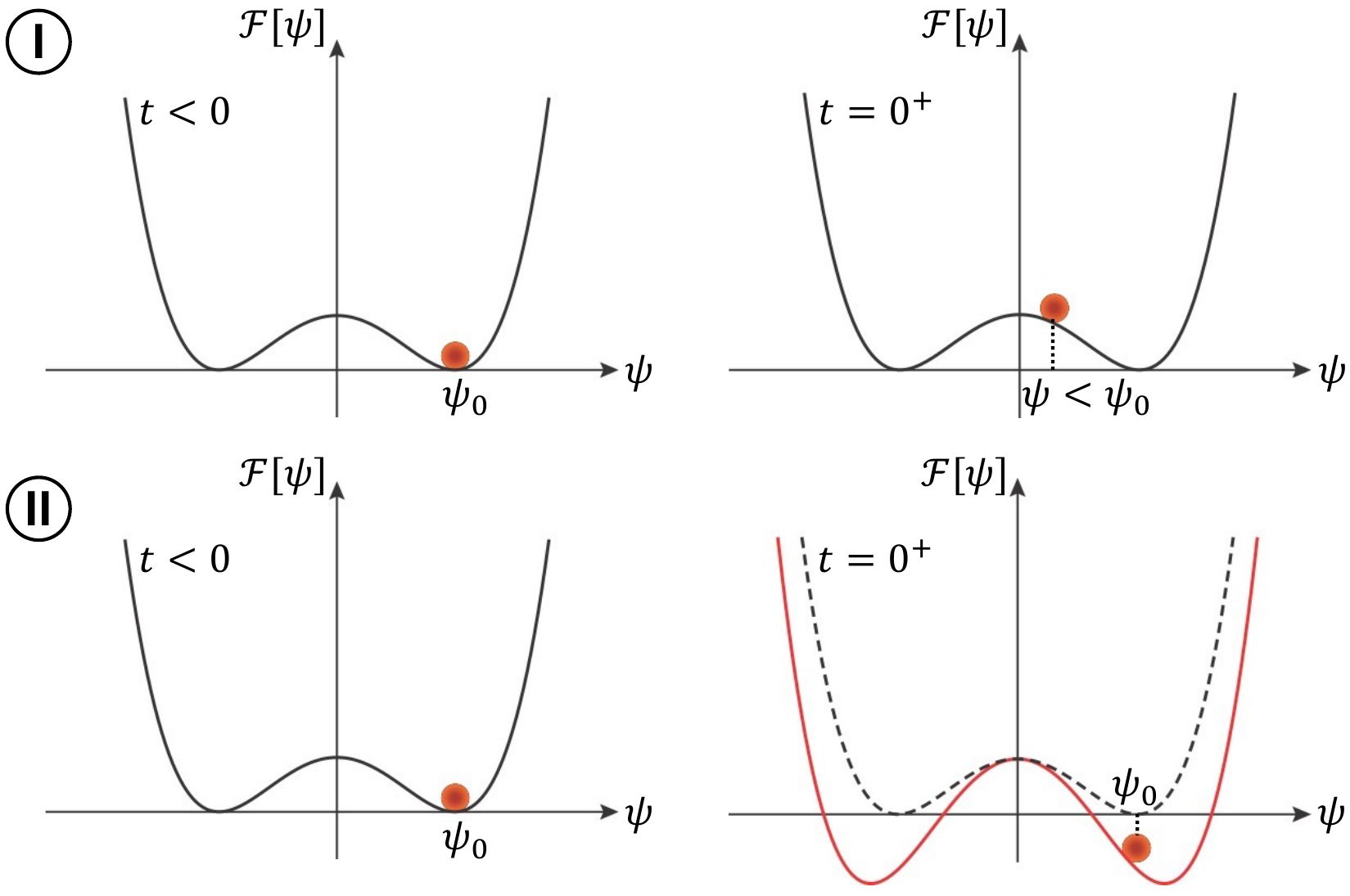}
	\caption{Two ways of modeling an impulsive optical quench: the laser pulse partially suppresses the order parameter expectation value (I) or it affects the coefficients of the Ginzburg-Landau free energy (II). 
	}
\label{fig:LG}
\end{figure}

\subsection{Quenching protocol}

\begin{figure*}[t!]
	\centering
	\includegraphics[width=0.9\textwidth]{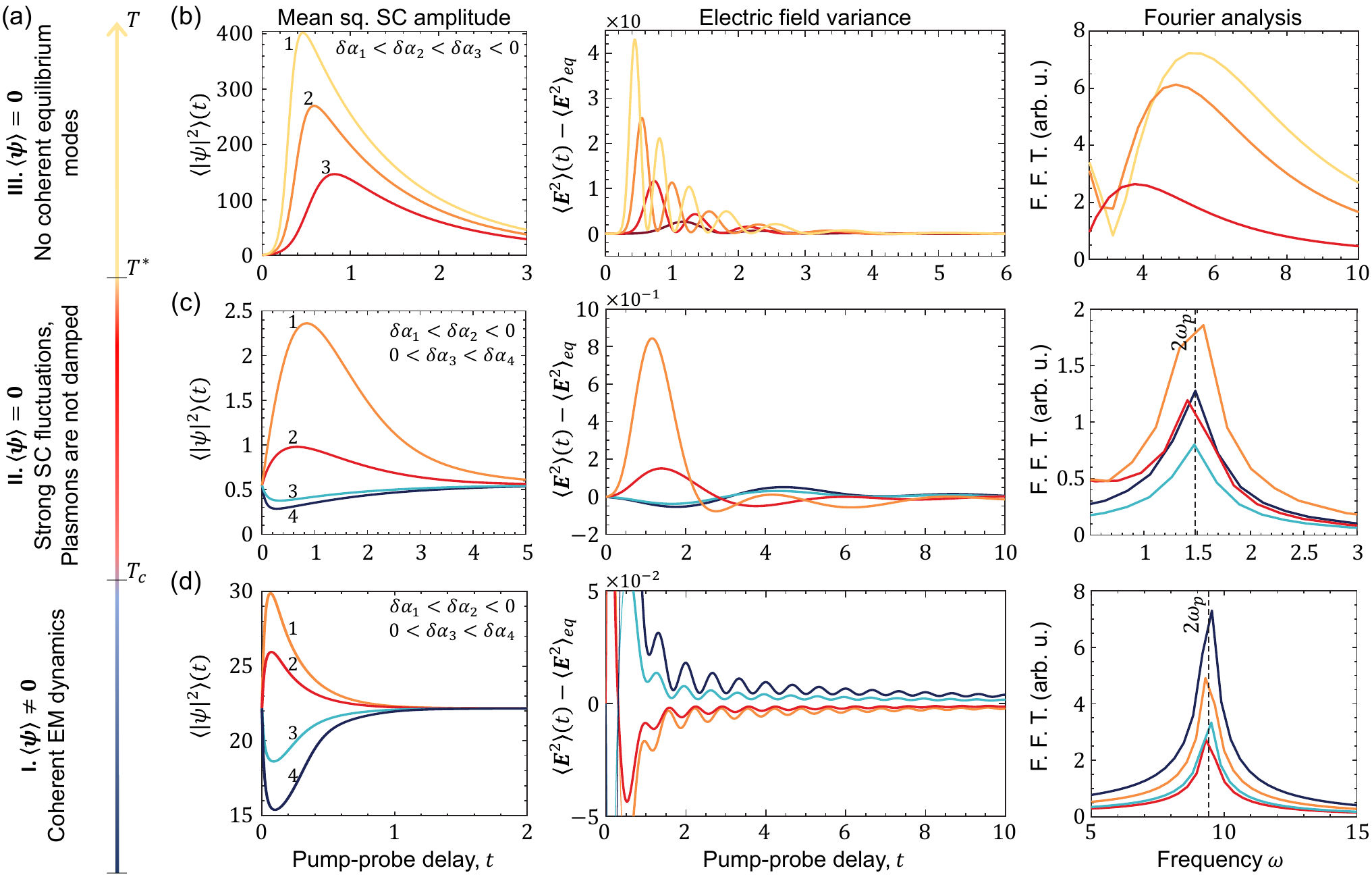}
	\caption{Post pulse dynamics in isotropic superconductors. (a) Schematic phase diagram. Depending on model parameters, there exist three regimes: i) for $T\leq T_c$, the long-range order parameter expectation value is nonzero $\langle \psi \rangle \neq 0$, the plasmon gap is notable, and plasmon dynamics is coherent; ii) $T_c<T \leq T^*$, no long-range order is present $\langle \psi \rangle = 0$, but superconducting equilibrium fluctuations, as captured by $\langle |\psi|^2 \rangle $, are sufficiently strong, and the correlation length is sufficiently long so that plasmons are still essentially coherent excitations; iii) $T^* < T$, plasmons become overdamped. Quenches in I are shown in lower panels (d): the order parameter evolution, encoded in $\langle |\psi|^2 \rangle(t)$, looks overdamped; in contrast, for $t\gtrsim 1$, we find that the electromagnetic field, as captured by $\langle \bm E^2\rangle(t) - \langle \bm E^2\rangle_{eq}$, shows periodic dynamics. The frequency of the oscillations, regardless of photosuppression or photostimulation of superconductivity,  equals twice the plasmon equilibrium frequency $2\omega_p$, as shown in the right panel, where we performed F.F.T. of the tails of the curves in the middle panel, with an exponential fit subtracted. The stronger the photoexcitation, the larger the amplitude of oscillations. 
	Quenches in II are shown in the middle panels (c). The post-pulse evolution here is similar to quenches in I, since plasmons still represent coherent excitations in equilibrium due to the strong superconducting fluctuations. (b) Interestingly, provided the order parameter relaxation rate is small, one can get periodic dynamics in III via strong photo-stimulation of superconductivity: the post-pulse dynamics of the electromagnetic field displays oscillations, though with poorly defined frequency. This is because $\langle |\psi|^2\rangle (t)$, the quantity that defines the plasmon frequency, is changing in time. Besides this, the quasiparticle conductivity is large, which makes the oscillations to be damped, cf. Eq.~\eqref{eqn_plasmon}. Parameters used: $\tau = 100$, $\tau_E = 5$, $\tau_{\alpha} = 1$, $\chi^{-1} = 0.1$, $\kappa = 5$, $\sigma = 1$, $Tr_0 = 10^{-2}$ (upper panels); $\tau = 5$, $\tau_E = 1$, $\tau_{\alpha} = 1$, $\chi^{-1} = 0.1$, $\kappa = 5$, $\sigma = 0.1$, $Tr_0 = 10^{-2}$ (middle panels); $\tau = 1$, $\tau_E = 1$, $\tau_{\alpha} = 0.2$, $\chi^{-1} = 0.1$, $\kappa = 25$, $\sigma = 0.1$, $Tr_0 = 10^{-2}$ (lower panels) -- see Appendix~\ref{appendix_units} for more details.
	}
\label{fig:Quenches}
\end{figure*}

We turn to discuss an impulsive optical quench. The effect of such incoherent photoexcitation is either to suddenly partially suppress the superconducting amplitude (Scenario~I in Fig.~\ref{fig:LG}) or to promptly perturb the superconducting free energy potential (Scenario~II in Fig.~\ref{fig:LG}). Scenario I captures quick condensate depletion as the laser pulse melts some of the Cooper pairs. In this scenario, the order parameter displays abrupt dynamics during the pump pulse but then evolves slowly on the time scale controlled by the TDGL relaxation time $\tau$, cf. Eq.~\eqref{eqn:TDGL}. Scenario II neglects partial condensate evaporation and describes photoexcitation as a sudden change to the coefficients of the Ginzburg-Landau free energy. The order parameter reacts relatively smoothly to this perturbation. In real materials, both scenarios are expected to play a role. Our numerical analysis of photoexcitation dynamics indicates that they give similar post-pulse phenomenology. For concreteness, throughout the rest of the paper, we primarily focus on the Scenario~II [see Appendix~\ref{appendix:Scenario I} for further discussion of Scenario~I]. Specifically, we consider quenches in the quadratic coefficient $\alpha(t)$ of the free energy of the form:
\begin{align}
    \alpha(t) = \alpha_0 + \delta \alpha \exp(-t/\tau_{\alpha}) \theta(t).\label{eqn:quench}
\end{align}
Here $\alpha_0$ is the pre-pulse value. $\delta \alpha$ encodes the strength of the photoexcitation. The case $\delta \alpha > 0$ describes transient suppression of superconductivity and could be due to laser-heating of quasiparticles. The situation with $\delta \alpha < 0$ represents transiently enhanced superconductivity and could arise in superconductors with competing orders, such as spin or charge density waves, which become suppressed by the photoexcitation. $\theta(t)$ is the Heaviside step function. $\tau_{\alpha}$ is a phenomenological relaxation rate.  One can easily generalize the present framework to various possible quenching protocols, including quenches in the sample temperature or modification of the dynamics in $\alpha(t)$ (for instance, use the Rothward-Taylor approach~\cite{PhysRevLett.19.27}). Regardless of the specific impulsive quenching protocol, our conclusions are insensitive to the functional form in Eq.~\eqref{eqn:quench}.

We remark that, in case the laser frequency is comparable to those of the relevant collective modes, one might be interested in quenches of the external electromagnetic field that directly affects order parameter dynamics~\cite{PhysRevB.100.104521}. We leave the study of these quenches to future work.

\section{Results and discussion}

We first investigate photoexcitation dynamics in the symmetry-broken phase in isotropic superconductors. The summary of our results is shown in Fig.~\ref{fig:Quenches}(d). Let us discuss photo-enhancement of superconductivity first, corresponding to quenches with $\delta \alpha < 0$, cf. Eq.~\eqref{eqn:quench}. In response to such a pump pulse, $\langle |\psi|^2 \rangle (t)$ is first transiently enhanced ($t\lesssim 0.2$) and then quickly returns to its equilibrium value (see Appendix~\ref{appendix:broken_phase} for more details). The stronger the photoexcitation, the stronger this quantity develops. We also find that the recovery is exponential, in contrast to the slow power-law dynamics in incommensurate CDWs~\cite{PhysRevB.101.174306}. The character of the recovery is determined by the equilibrium collective modes: For superconductors, the plasmons are gapped, resulting in a quick exponential recovery, while it is gapless phasons (Goldstone modes) that are responsible for the slow evolution in CDWs. Now we turn to the dynamics of the electromagnetic field, which shows additional features. While the ensemble averaged electric field in the sample remains zero $\langle \bm E\rangle (t) = 0$ (a homogeneous quench cannot result in the development of nonzero electric field since fluctuations with $\pm \bm E$ are equally likely), the evolution of electromagnetic fluctuations, as captured by $\langle \bm E^2\rangle (t) - \langle \bm E^2\rangle_{eq}$, is governed by three stages: i) initial enhancement, signaling plasmon proliferation; ii) transient suppression below the equilibrium value; and iii) recovery. The second stage can be understood as follows. Since at longer times $\langle |\psi|^2 \rangle(t)$ exceeds its equilibrium value and since this quantity determines the plasmon frequency at equilibrium, it renders the plasmons to be energetically costly, resulting in their eventual depopulation. Most remarkably, as $\langle \bm E^2\rangle (t)$ recovers, it acts a Raman mode and oscillates with a frequency twice the plasmon gap $2\omega_p$ [Eq.~\eqref{eqn_plasmon} with $\gamma = 1$]. We note that for the case of photo-suppression of superconductivity, corresponding to quenches with $\delta \alpha > 0$, the phenomenology of the transient dynamics is ``flipped'' compared to the case of photo-enhancement, but oscillations are still present.

Quenches in the pseudogap phase are shown in Fig.~\ref{fig:Quenches}(c), and we find that the phenomenology of transient dynamics is similar to that in the symmetry broken phase. There are two qualitative differences: the plasmon frequency in the pseudogap phase might be notably smaller, and the plasmon lifetime is shorter since $\sigma$ increases with increasing temperature. Above $T_c$, the order parameter expectation value $\langle \psi\rangle = 0$ remains zero, even for quenches into the symmetry broken phase. We conclude that in order to gain oscillatory dynamics, it is sufficient to have appreciable superconducting fluctuations rather than long-range coherence.

\begin{figure}[t!]
	\centering
	\includegraphics[width=1\linewidth]{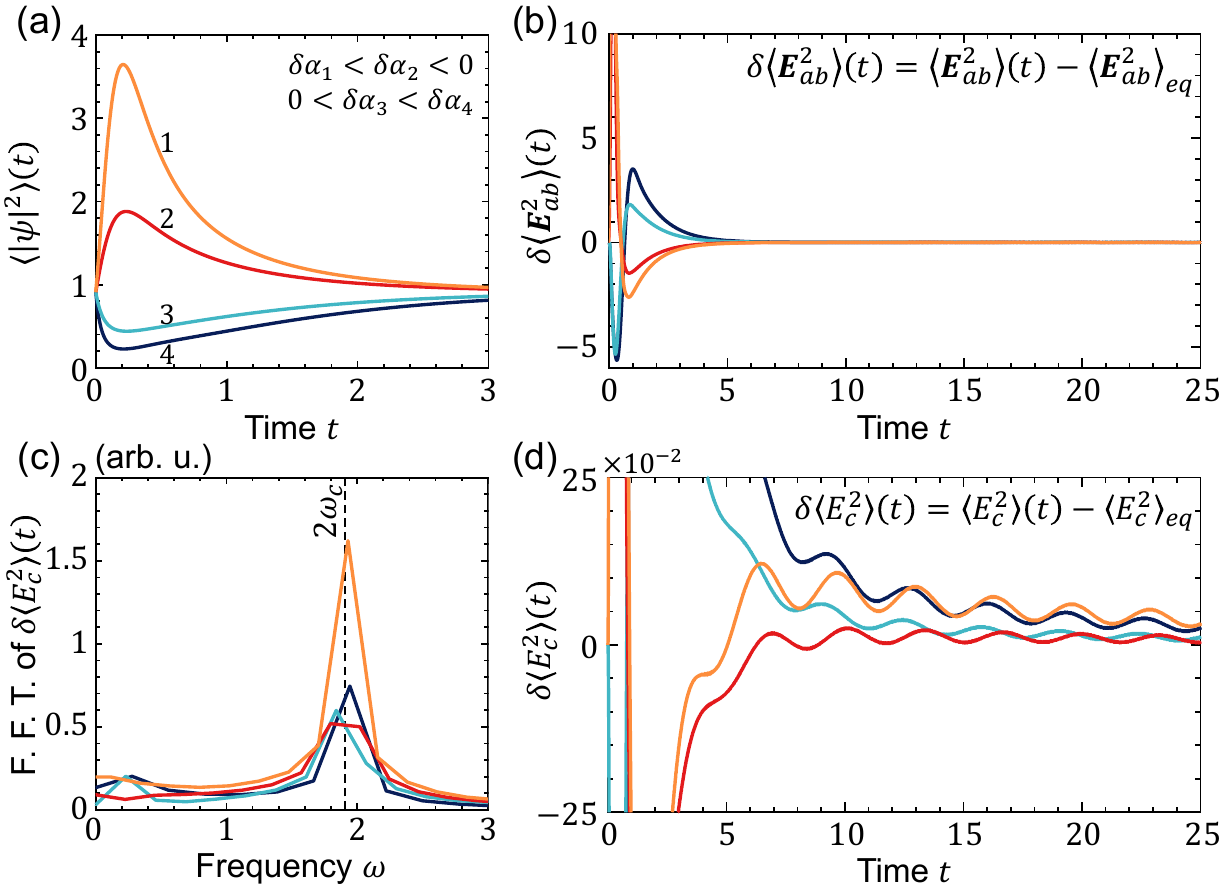}
	\caption{Photoexcitation dynamics in anisotropic superconductors in the pseudogap phase above $T_c$. (a) Evolution of $\langle |\psi|^2 \rangle (t)$ looks overdamped regardless of the quenching conditions.
	(b) Similarly, the in-plane electric field also displays overdamped dynamics due to the large in-plane plasmon gap $\omega_{ab} \gg \omega_c$.
	(c and d) In contrast, the out-of-plane electric field variance exhibits oscillations with frequency $2\omega_c$ (d), as shown in F.F.T. analysis in (c). These oscillations are long-lived because the out-of-plane normal conductivity $\sigma_c$ is small. Parameters used: $\tau = 5$, $\tau_E = 10^{-2}$, $\tau_{\alpha} = 0.1$, $\chi^{-1} = 0$, $\kappa = 10^3$, $\sigma_{ab} = 0.1$, $\sigma_{c} = 10^{-5}$, $Tr_0 = 10^{-3}$, %$\alpha_0 = -1$,
	$\gamma = 10$.
	}
\label{fig:Anisotropic}
\end{figure}

Since there are no coherent collective modes at equilibrium for $T>T^*$, it would be compelling if one could induce periodic dynamics via photoexcitation, i.e., by putting the system out-of-equilibrium. Such a situation might indeed occur provided the order parameter relaxation rate is small [see Appendix~\ref{appendix:regime_3} for additional discussion], as shown in Fig.~\ref{fig:Quenches}(b). The possibility of achieving oscillatory evolution with strong photoexcitation, in an otherwise incoherent system, warrants further experimental studies.

Of particular experimental interest are cuprates, layered superconductors with the $c$-axis Josephson plasmon gap being in the terahertz range and $\gamma \gg 1$. Furthermore, these materials have small out-of-plane quasiparticle conductivity, which renders the $c$-axis plasmons long-lived. Figure~\ref{fig:Anisotropic} shows our results for quenches in anisotropic superconductors. Similarly to the isotropic situation, the order parameter dynamics remains overdamped, regardless of the quenching protocol [Fig.~\ref{fig:Anisotropic}(a)]. We also find that the in-plane electric field displays no signatures of coherent evolution [Fig.~\ref{fig:Anisotropic}(b)] (we recall that the in-plane plasmon gap is a large energy scale). However, the out-of-plane electric field, as captured by $\langle E_c^2\rangle (t) - \langle E_c^2\rangle_{eq}$, demonstrates lasting oscillatory dynamics [Fig.~\ref{fig:Anisotropic}(d)], with twice the $c$-axis Josephson plasmon frequency [Fig.~\ref{fig:Anisotropic}(c)], $2\omega_c$. Since it is natural to have small out-of-plane conductivity in anisotropic materials, we conclude that cuprates are promising materials to test our findings experimentally.

\begin{figure}[t!]
	\centering
	\includegraphics[width=1\linewidth]{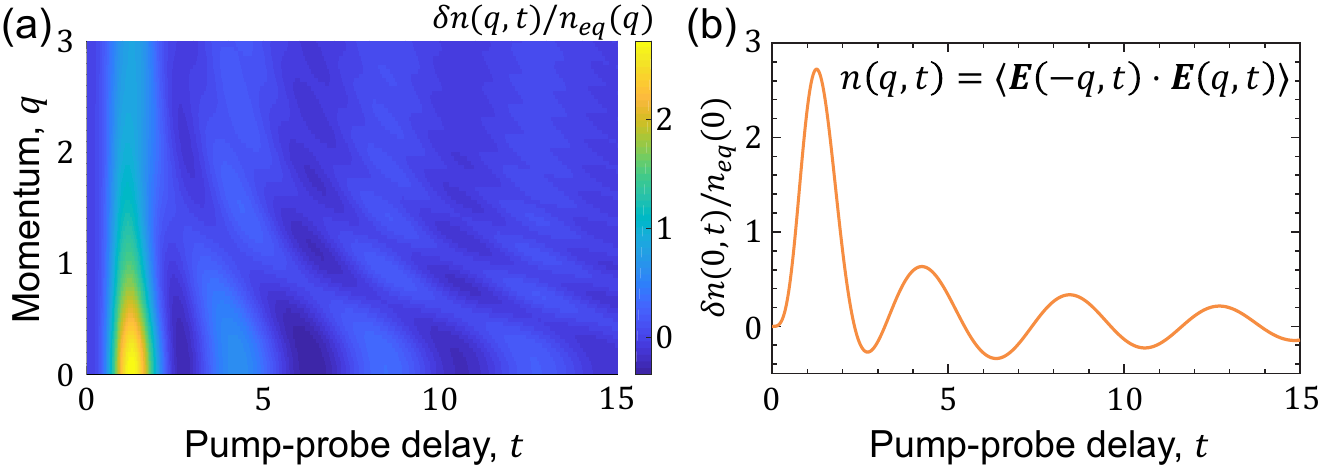}
	\caption{ Post-pulse dynamics of individual momentum harmonics. (a) Evolution of $\delta n(q,t) = n(q,t) - n_{eq}(q)$, where $n(q,t) = \langle \bm E(-q,t) \cdot \bm E(q,t)\rangle$, illustrates that the entire plasmon continuum gets excited. Low momenta modes are affected more strongly compared to large momenta ones. (b) Dynamics of $\delta n(q,t)$ at $q = 0$ displays the bi-plasmon periodic dynamics. For typical model parameters, the relative amplitude of oscillating fluctuations can be $\sim 100\%$ of the thermal equilibrium fluctuations $n_{eq}(0)$. Remarkably, during these oscillations, the thermal noise can be reduced by $\sim 40\%$ at the bottom of the first cycle, which is clear evidence for the vacuum squeezing. Parameters are the same as in Fig.~\ref{fig:Quenches}(c).
	}
\label{fig:Fig4}
\end{figure}

%We turn to discuss the role of the quenching protocol, which affects both the amplitude and lifetime of the bi-plasmon oscillations. 
For experimental detection of these coherent dynamics, it is important to characterize the amplitude and the lifetime of the bi-plasmon oscillations, which can be affected by the quenching conditions. A stronger laser pulse, as encoded in $|\delta \alpha|$, would lead to a greater number of low momenta plasmons. As these plasmons are responsible for the periodic dynamics, there would be a larger amplitude of the oscillatory dynamics. Our simulations indicate that this amplitude grows linearly with the pulse strength. In isotropic superconductors, where the separation between longitudinal and transverse modes is possible, the former are excited stronger and constitute the primary source of temporal oscillations of the electric field variance. In anisotropic materials, such separation is not possible away from high symmetry directions, and all modes contribute to the bi-plasmon oscillations. In such systems, a different consideration is crucial for understanding which JPs are excited by the pulse: modes with larger in-plane currents have stronger damping due to the large anisotropy of the normal fluid conductivity. Dynamics of individual momentum components, $n(q,t) = \langle \bm E(-q,t)\cdot \bm E(q,t)\rangle$, show that the strength of their oscillations, for example at $q = 0$, can even exceed the value of equilibrium fluctuations $n_{eq}(q=0)$ [Fig.~\ref{fig:Fig4}(b)]. 

The primary decay mechanism of oscillations is dephasing, which depends on details of the quenching protocol, including pulse duration and pulse frequency profile, as well as on the plasmon spectrum. During photoexcitation, the entire plasmon continuum is excited, with high-energy plasmons getting less populated compared to the low-energy ones [Fig.~\ref{fig:Fig4}(a)]. Since essentially all different frequencies above twice the plasmon gap contribute to the evolution of the electromagnetic energy density, these different modes dephase and render the oscillations damped. %For smoother order parameter dynamics, this effect is weaker, resulting in longer oscillations. 
Dephasing is weaker and oscillations last longer for plasmons with flatter dispersion because, in this case, the impulsive photoexcitation produces plasmon pairs with frequencies that are closer to each other in energy [see Appendix~\ref{appendix:Scenario I} for additional discussion]. The dispersion of the relevant plasmons is expected to be flatter in type-II superconductors with a large Ginzburg parameter (see Appendix~\ref{sec: Collective modes}), such as in cuprates.

As a final remark, the mechanism reported in this paper for converting incoherent pumping to long-lived oscillations relies only on the existence of a gapped phase mode in a $U(1)$ symmetry-broken state. As such, in any scenario where such a gapped mode exists, we expect that any perturbation of the corresponding underlying order parameter, irrespective of transient enhancement or suppression, will lead to similar Raman-like oscillations. Potential examples include the relative phase mode of an unconventional superconductor~\cite{Poniatowski.2021,Zeng.2021}, structural phonon mode in a crystal~\cite{Juraschek.2020}, or the pseudo-Goldstone mode thought to exist in the putative excitonic insulator $\rm Ta_2 NiSe_5 $ \cite{Baldini20}. In the latter, direct coupling of the excitonic condensate to phonons breaks the $U(1)$ symmetry, explicitly making the Goldstone mode gapped.

\section{Conclusion}
Here we demonstrated how generic quenches of the superconducting order parameter result in oscillatory dynamics at frequency twice the plasma resonance. These oscillations are made up of pairs of plasma fluctuations and can directly act as a parametric drive to photons. As a consequence, experimental observables such as reflectivity and conductivity will be dynamically renormalized, as discussed in Refs.~\cite{marios2020,vanHoegen19,Kaiser14,Hu14} on Floquet driven superconductors. In particular, in optical reflectivity, one expects the development of resonance at half the oscillation frequency, i.e., at the plasmon gap. Besides this, the parametric drive can result in amplification of reflectivity to values exceeding unity for large driving amplitudes~\cite{Buzzi21}. This mode can manifest as a blue-shift of the effective plasma resonance in conductivity, as observed in Ref.~\cite{Okamoto2016} for a bilayer superconductor. Apart from consequences for optical reflectivity, oscillations predicted here act as a Raman active mode and, thus, should be detectable as a modulation in the time-dependent optical properties. Our work implies that the presence of special collective modes, such as the Higgs-amplitude excitation or phonons, is not a prerequisite for providing a parametric drive in photoexcited superconductors. Importantly, our results are generic to a wide variety of superconductors, paving the way for future experimental discovery of 
Floquet quantum matter with unusual optical properties.

%  For theoretical outlook, it seems interesting to extend the presented framework to include coherent order parameter dynamics, where the Higgs mode is no longer overdamped. Such a situation might be particularly interesting in fullerides at temperatures well below $T_c$, where one can study potentially rich interplay between the Higgs oscillations and the bi-plasmon oscillations predicted here. Another promising direction is to extend our framework to study quenches in the electric field that is directly coupled to order parameter. Such situ

\section*{ACKNOWLEDGEMENTS}
The authors would like to thank M. Mitrano, L. Glazman, I. Klich, B. Halperin, D. Nicoletti, A. von Hoegen, M. Fechner, M. F\"{o}rst, P. Narang, I. Esterlis, and S. Chatterjee for stimulating discussions. P.E.D, M.H.M., and E.D. were supported by Harvard-MIT CUA, AFOSR-MURI: Photonic Quantum Matter award FA95501610323, Harvard Quantum Initiative. A.Z. acknowledges support from the Miller Institute for Basic Research in Science. J.B.C. is an HQI Prize Postdoctoral Fellow and gratefully acknowledges support from the Harvard Quantum Initiative. D.P. acknowledges financial support by the Israel Science Foundation (Grant No. 1803/18). A.C.'s work was funded in part by the European Research Council under the European Union's Seventh Framework Programme (FP7/2007–2013)/ERC (grant agreement no. 319286 (QMAC)) and by the Cluster of Excellence `CUI: Advanced Imaging of Matter' of the Deutsche Forschungsgemeinschaft (DFG), EXC 2056, project ID 390715994.

\section*{AUTHOR CONTRIBUTION}
E.D. and A.C. supervised the project. P.E.D. developed the theoretical framework. All authors contributed to the development of main ideas and results. P.E.D, A.Z., M.H.M., and E.D. wrote the manuscript with substantial input from all the authors.

\bibliography{SC}

\onecolumngrid
\appendix

\section{Dimensionless units}
\label{appendix_units}

At equilibrium, it is convenient to define dimensionless quantities as, cf. Ref.~\cite{PhysRevB.46.8376}:
\begin{gather}
    {\bm r} =  \lambda_{ab} \tilde{\bm r},\quad t = \frac{\tilde{t}}{|\alpha|},\quad  \psi = \sqrt{ \frac{|\alpha|}{\beta}} \tilde{\psi},\quad
    {\bm A} = \sqrt{2} \lambda_{ab} H_c \tilde{\bm A},\\
     {\bm E} = \frac{ \sqrt{2} \lambda_{ab} H_c  |\alpha|}{c} \tilde{\bm E},\quad
    \rho = \frac{\sqrt{2}|\alpha| H_c}{4\pi c}\tilde{\rho},\quad \chi = \frac{\sqrt{2} H_c }{4\pi c}\tilde{\chi},\quad \sigma = 
    \frac{c^2}{4\pi \lambda_{ab}^2 |\alpha| } \tilde{\sigma},
    \label{eqn:dim_q}
\end{gather}
where  $\lambda_{ab} = \sqrt{m_{ab} c^2 \beta/4\pi (e^*)^2 |\alpha|}$ defines the unit of length. $\lambda_{c} = \gamma \lambda_{ab}$ is the magnetic penetration depth for currents along the $z$-axis. $\xi_{ab} = \sqrt{1/2m_{ab} |\alpha|}$ and $\xi_{c} = \sqrt{1/2m_{c} |\alpha|}$ are the two coherence lengths. The Ginzburg parameter is:
\begin{align}
   \kappa = \frac{\lambda_{ab}}{\xi_c}= \frac{\lambda_{c}}{\xi_{ab}}.
\end{align}
$H_c = \sqrt{4\pi \alpha^2/\beta}$ is the thermodynamic critical field. We also introduce $r_0 = 2\pi (e^*)^2/\lambda_{ab} m_{ab} c^2$, which is nothing but the classical Cooper pair radius in the unit of $\lambda_{ab}$. 

When it comes to dynamics, here we are primarily interested in quenches in the Landau coefficient $\alpha(t)$. We therefore fix all the dimensionless quantities in Eq.~\eqref{eqn:dim_q} at some reference $\alpha_0$ (for example, it can be chosen to be the $T=0$ value $\alpha(T=0)$ or the pre-pulse equilibrium value) and write the equations of motion in the dimensionless units as (for notational simplicity, here and below, we drop the tilde sign for dimensionless quantities):
\begin{gather}
    \tau (\partial_t + i\chi^{-1}\nabla \cdot \bm E) \psi = - (-i\gamma\kappa^{-1}\nabla_{ab} - \bm A_{ab})^2\psi - \gamma^{-2}(-i\gamma\kappa^{-1}\partial_{z} -  A_z)^2\psi\label{eqn:psi_t}
    -\alpha\psi - |\psi|^2\psi + \eta,\\
    \partial_t \bm A = - \bm E,\\
    \tau_E \partial_t\bm E = \nabla\times\nabla\times\bm A - \hat{\sigma} (\bm E- \gamma\kappa^{-1}\chi^{-1}\nabla (\nabla\cdot\bm E))
    - \frac{1}{2} \hat{m}^{-1}(\psi^*(-i\gamma\kappa^{-1}\nabla - \bm A)\psi + c.c.) + \bm \xi, \label{eqn:E_t}
\end{gather}
where $\hat{m}^{-1} = \text{diag}(1,1,\gamma^{-2})$ reflects the anisotropy of the superconductor. The thermal noise terms in the Lagnevin equations~\eqref{eqn:psi_t} and~\eqref{eqn:E_t} obey: 
\begin{gather}
    \langle \eta^*({\bm r},t) \eta({\bm r}',t') \rangle = 4 T \tau r_0 \delta( {\bm r - \bm r'})\delta(t-t'),\\
    \langle\xi_\alpha({\bm r},t) \xi_\beta({\bm r}',t') \rangle  = 2 T\sigma_{\alpha \beta}  r_0 \delta( {\bm r - \bm r'})\delta(t-t'),
\end{gather}
such that the fluctuation-dissipation theorem is satisfied~\cite{kamenev2011field}. Here $\tau_E = \lambda_{ab}^2\alpha^2/c^2$. Below we define $\Gamma_E = \tau_E^{-1}$.

\section{Mean-field analysis of equilibrium collective modes}
\label{sec: Collective modes}

We turn to discuss collective equilibrium excitations in the symmetry broken phase. To this end, we neglect the noise terms, linearize the above equations on top of $\psi =  {\bar \psi} (1 + i\theta) + \delta \Delta$ (${\bar \psi}  = \sqrt{-\alpha}$), and obtain the spectrum of collective modes. We find that even though the dynamics of the order parameter amplitude $ \delta \Delta$ is overdamped, the dynamics of the phase $\theta$ is not, and therefore, we focus on these modes. We consider the cases of isotropic and anisotropic superconductors separately.

\subsection{Isotropic superconductors}

The linearized supercurrent reads: ${\bm j}_s \approx \bar{\psi}^2(\kappa^{-1} \nabla \theta - \bm A)$. We decompose all vectors in momentum space into the transverse and longitudinal components: ${\bm A}({\bm k},\omega) = {\bm A}_\perp({\bm k},\omega) + {\bm A}_\parallel({\bm k},\omega)$, where ${\bm A}_\parallel$ points along $\bm k$ and ${\bm A}_\perp$ is orthogonal to $\bm k$. We find that the transverse sector decouples from the rest of the system, and its spectrum is given by:
\begin{align}
    \omega_\perp(k) = \sqrt{(k^2 + {\bar \psi}^2)\Gamma_E - \frac{1}{4} \sigma^2 \Gamma_E^2} -\frac{i}{2} \sigma \Gamma_E,
\end{align}
i.e. the non-zero expectation value ${\bar \psi}$ opens up the plasmon gap, in accordance with the Anderson-Higgs mechanism and the Meissner effect. The primary role of the normal conductivity is to provide damping and redshift these transverse plasmon excitations.

Longitudinal waves are coupled to the dynamics of the order-parameter phase $\theta$:
\begin{gather}
    \Big[\tau_E \omega^2 + i\omega \sigma \Big( 1 + \frac{k^2}{\kappa\chi}\Big) - \bar{\psi}^2\Big]A_\parallel + \frac{i k\bar{\psi}^2 \theta}{\kappa} = 0,\\
    \Big[-i\omega \tau + \frac{k^2}{\kappa^2} \Big]\theta - \Big[\frac{\omega \tau}{\chi} - \frac{i}{\kappa} \Big] k A_\parallel = 0.
\end{gather}
By solving this system analytically, we obtain the following quadratic equation defining the spectrum of the longitudinal modes:
\begin{align}
    \tau_E\omega^2 + i\omega \Big[ \sigma  \Big(1 + & \frac{k^2}{\kappa\chi}\Big)  + \tau_E \Gamma\frac{k^2}{\kappa^2} \Big]  - \Big(1 + \frac{k^2}{\kappa\chi}\Big)\Big[\bar{\psi}^2 + \sigma\Gamma \frac{k^2}{\kappa^2}\Big] = 0.
\end{align}
Similarly to the transverse sector, the longitudinal modes also open up the same plasmon gap. Because of the Coulomb screening and the coupling to the order parameter phase, the longitudinal excitations are more damped for $k\neq 0$ than the transverse ones.

\subsection{Anisotropic superconductors}
The linearized supercurrent modifies to
\begin{align*}
    {\bm j}_s \approx \bar{\psi}^2\Big(\frac{\gamma}{\kappa}\nabla_{ab}\theta - \bm A_{ab}\Big) + \frac{\hat{z}\bar{\psi}^2}{\gamma^2}\Big(\frac{\gamma}{\kappa}\partial_z \theta - A_z\Big).
\end{align*}
Due to the anisontropy, we now decompose all vectors as: $\bm A (\bm k) = A_\parallel \hat{\bm k}_{ab} + A_\perp \hat{\bm k}_{ab} \times \hat{\bm z} + A_z\hat{\bm z}$, etc. Here $\bm k = (\bm k_{ab}, k_z)$. We find that the transverse sector (with vectors pointing along $\hat{\bm k}_{ab} \times \hat{\bm z}$) decouples from the rest of the system, and its spectrum is given by:
\begin{align}
\omega_\perp(\bm k) = \sqrt{(k^2 +\bar{\psi}^2)\Gamma_E - \frac{1}{4}\sigma^2_{ab}\Gamma^2_E} -\frac{i}{2}\sigma_{ab} \Gamma_E.
\end{align}
As for the isotropic case, the transverse sector exhibits opening of the plasmon gap, expected to be a large energy scale.

The remaining linearized EM equations read:
\begin{gather}
    \tau_E \omega^2 A_\parallel =  k_z^2 A_\parallel - k_{ab} k_z A_z - {\bar \psi}^2\Big(\frac{\gamma}{\kappa}ik_{ab}\theta - A_\parallel\Big) - i\omega \sigma_{ab}\Big(A_\parallel + \frac{\gamma}{\kappa\chi} k_{ab}(k_{ab} A_\parallel + k_z A_z) \Big), \label{eqn:An_Apar_cm}\\
    \tau_E \omega^2 A_z =  k_{ab}^2 A_z - k_{ab} k_z A_\parallel - \frac{{\bar \psi}^2}{\gamma^2}\Big(\frac{\gamma}{\kappa}ik_z\theta - A_z\Big)  - i\omega\sigma_{c}\Big(A_z + \frac{\gamma}{\kappa\chi} k_z(k_{ab} A_\parallel + k_z A_z)\Big).\label{eqn:An_Az_cm}
\end{gather}
Linearized equation for the order parameter phase $\theta$ is:
\begin{align}
    \Big[-i\omega \tau + \frac{\gamma^2 k_{ab}^2}{\kappa^2} + & \frac{k_{z}^2}{\kappa^2}  \Big]\theta - \frac{\omega \tau}{\chi} (k_{ab} A_{\parallel} + k_z A_z )  + \frac{i\gamma}{\kappa}  (k_{ab} A_{\parallel} + \gamma^{-2}k_z A_z ) = 0.\label{eqn:theta_an_cm}
\end{align}
To find the spectrum of collective modes, one can solve Eqs.~\eqref{eqn:An_Apar_cm}-\eqref{eqn:theta_an_cm} numerically. Here we are mostly interested in the limit $\bm k \to 0$:
\begin{gather}
\omega_\parallel(\bm k = 0) = \sqrt{\bar{\psi}^2\Gamma_E  - \frac{1}{4}\sigma^2_{ab}\Gamma^2_E} -\frac{i}{2}\sigma_{ab} \Gamma_E ,\\
\omega_z(\bm k = 0) =  \sqrt{\frac{\bar{\psi}^2}{\gamma^2} \Gamma_E - \frac{1}{4}\sigma^2_{c}\Gamma^2_E} -\frac{i}{2}\sigma_{c} \Gamma_E.
\end{gather}
We conclude that all of the collective modes are gapped. Most importantly, though, we find that the $c$-axis gap is factor of $\gamma$ smaller than the gap of the other two plasmons (for a related discussion in layered superconductors, see Ref.~\cite{PhysRevB.50.12831}). In anisotropic superconductors, such as cuprates, this gap is in the terahertz range, rendering the $c$-axis Josephson plasmons to be the primary low-energy excitations. We also note that their lifetime is large, since the out-of-plane conductivity $\sigma_c$ is small.

\section{Equations of motion within the Gaussian approximation}
\label{Sec:Main_eqn}

%In the presence of a homogeneous time-dependent external electric field, one %We further will assume that $\bm E_{\rm ext}$ points along the $z$-axis.
The equations of motion in the dimensionless units and in momentum space read:
\begin{align}
    \tau \partial_t \psi_1({\bm k}, t) &  = i \tau\chi^{-1} \int\limits_{\bm q} \bm q \cdot \bm E(\bm q,t) \psi_2({\bm k} - \bm q,t) - ( \alpha(t) + \gamma^2\kappa^{-2}k_{ab}^2 + \kappa^{-2}k_z^2) \psi_1({\bm k},t)  \notag\\
    & + i \gamma \kappa^{-1} \int\limits_{\bm p} ({\bm k} + {\bm p}) \hat{m}^{-1} {\bm A}({\bm k} - {\bm p},t) \psi_2({\bm p},t)  \notag\\
    & - \int\limits_{\bm p_1,\bm p_2} [\bm A({\bm p}_1,t) \hat{m}^{-1} \bm A(\bm p_2,t) + \psi_1({\bm p_1},t)\psi_1(\bm p_2,t) +\psi_2(\bm p_1,t)\psi_2(\bm p_2,t) ] \psi_1({\bm k - \bm p_1 - \bm p_2},t) + \eta_1({\bm k},t), \label{eqn:SI_psi1_stochastic}\\
   \tau \partial_t \psi_2({\bm k}, t) &  = -i \tau\chi^{-1} \int\limits_{\bm q} \bm q \cdot \bm E(\bm q,t) \psi_1({\bm k} - \bm q,t) - ( \alpha(t) + \gamma^2\kappa^{-2}k_{ab}^2 + \kappa^{-2}k_z^2) \psi_2({\bm k},t) \notag\\
   & - i\gamma \kappa^{-1} \int\limits_{\bm p} ({\bm k} + {\bm p})\hat{m}^{-1} {\bm A}({\bm k} - {\bm p},t) \psi_1({\bm p},t) \notag\\
    &  - \int\limits_{\bm p_1,\bm p_2} [\bm A({\bm p}_1,t) \hat{m}^{-1} \bm A(\bm p_2,t)  + \psi_1({\bm p_1},t)\psi_1(\bm p_2,t) +\psi_2(\bm p_1,t)\psi_2(\bm p_2,t) ] \psi_2({\bm k - \bm p_1 - \bm p_2},t) + \eta_2({\bm k},t),\\
    \partial_t \bm  A({\bm k}, t) &  = -\bm E(\bm k,t),\\
    \tau_E \partial_t \bm  E({\bm k}, t) &  = k^2 \bm A(\bm k,t) - (\bm k \cdot \bm A({\bm k},t))\bm k - \hat{\sigma}_{\alpha\beta} ( \delta_{\beta \gamma}+ \gamma\kappa^{-1}\chi^{-1}k_\beta k_\gamma) E_\gamma(\bm k,t) - \bm j_s(\bm k,t) + \bm \xi(\bm k,t ),\label{eqn:SI_E_stochastic}
\end{align}
where $\psi_1$ and $\psi_2$ are the real and imaginary components of the real-space order parameter $\psi(\bm r,t) = \psi_1(\bm r,t) + i\psi_2(\bm r,t)$.
$\langle \eta_a({\bm k},t) \eta_b({\bm k}',t') \rangle  = 2 T \tau r_0 \delta_{ab} \times (2\pi)^3 \delta({\bm k+\bm k}')\delta(t-t'),$ $\langle\xi_\alpha({\bm k},t) \xi_\beta({\bm k}',t') \rangle  = 2 \sigma_{\alpha \beta}  T r_0 \times (2\pi)^3 \delta({\bm k+\bm k}')\delta(t-t').$ Here $\int_{\bm q} \equiv \int \frac{d^3 \bm q}{(2\pi)^3}$. The superconducting current density in momentum space reads:
\begin{align}
    \bm j_s(\bm k,t) & = i \gamma \kappa^{-1} \hat{m}^{-1} \int\limits_{\bm p} (2{\bm p}-{\bm k}) \psi_1({\bm k}-{\bm p},t)\psi_2({\bm p},t)  \notag\\
    & - \int\limits_{\bm p_1,\bm p_2}  [\psi_1({\bm p}_1,t)\psi_1({\bm p}_2,t) + \psi_2({\bm p}_1,t)\psi_2({\bm p}_2,t)]\hat{m}^{-1}{\bm A}({\bm k}- {\bm p}_1-{\bm p}_2,t).
\end{align}

We note that Eqs.~\eqref{eqn:SI_psi1_stochastic}-\eqref{eqn:SI_E_stochastic} are stochastic first-order differential equations. They can be rewritten into a single first-order Fokker-Planck equation on the cumulative distribution functional ${\cal P}[t;\psi_1,\psi_2,\bm A,\bm E]$. Here we assume that ${\cal P}$ is a Gaussian distribution, and derive the corresponding equations on various correlation functions. For a related discussion in incommensurate charge density waves, see Ref.~\cite{PhysRevB.101.174306}. We also invoke translational symmetry. Specifically, below we introduce: $\psi_1(t) = \langle \psi_1(\bm k =0,t) \rangle$, ${\cal D}_{11}(\bm k, t) = \langle \psi_1(-\bm k, t) \psi_1(\bm k, t) \rangle_c$, ${\cal D}_{22}(\bm k, t) = \langle \psi_2(-\bm k, t) \psi_2(\bm k, t) \rangle_c$, $\pi_\alpha(\bm k, t) = \langle E_\alpha(-\bm k, t) \psi_2(\bm k, t) \rangle_c$, $a_\alpha(\bm k, t) = \langle A_\alpha(-\bm k, t) \psi_2(\bm k, t) \rangle_c$, $\Phi_{\alpha\beta}(\bm k, t) = \langle A_\alpha(-\bm k, t)A_\beta(\bm k, t)\rangle_c$, $K_{\alpha\beta}(\bm k, t) = \langle A_\alpha(-\bm k, t)E_\beta(\bm k, t)\rangle_c$, and $\Pi_{\alpha\beta}(\bm k, t) = \langle E_\alpha(-\bm k, t)E_\beta(\bm k, t)\rangle_c$. Other correlators: $\langle \psi_2(\bm k =0,t) \rangle$, $\langle \psi_1(-\bm k, t) \psi_2(\bm k, t) \rangle_c$,  $\langle E_\alpha(-\bm k, t) \psi_1(\bm k, t) \rangle_c$, $\langle A_\alpha(-\bm k, t) \psi_1(\bm k, t) \rangle_c$ -- turn out not to develop within the presented framework and, thus, can be neglected. We omit writing explicit dependence on time of the dynamical variables in the right-hand side of each of the equations below, unless it is needed.

{\textbf{SC sector}.} The order parameter dynamics and its fluctuations follow:
\begin{align}
     \tau\partial_t \psi_1(t)  & =  \int\limits_{\bm p} i\bm p \cdot (\gamma\kappa^{-1} \hat{m}^{-1}\bm a(\bm p) - \tau \chi^{-1} \bm \pi (\bm p)) - \psi_1 \Big( \alpha + \psi_1^2 +  \int\limits_{\bm p}( \text{tr}(\hat{m}^{-1}{\Phi}({\bm p})) + 3{\cal D}_{11}({\bm p}) + {\cal D}_{22}({\bm p})) \Big),\label{eqn:psi_1_full}\\
    \partial_t {\cal D}_{11}({\bm k},t)  & = 2 \Gamma T r_0 - 2 \Gamma {\cal D}_{11} \Big( \alpha + \gamma^2\kappa^{-2}k_{ab}^2 + \kappa^{-2}k_z^2 + 3 \psi_1^2 +  \int\limits_{\bm p}( \text{tr}(\hat{m}^{-1}{\Phi}({\bm p})) + 3{\cal D}_{11}({\bm p}) + {\cal D}_{22}({\bm p})) \Big),\\
     \partial_t {\cal D}_{22}({\bm k},t) & = 2 \Gamma T r_0 - 2 \Gamma {\cal D}_{22} \Big( \alpha + \gamma^2\kappa^{-2}k_{ab}^2 + \kappa^{-2}k_z^2 + \psi_1^2 +  \int\limits_{\bm p}( \text{tr}(\hat{m}^{-1}{\Phi}({\bm p})) + {\cal D}_{11}({\bm p}) + 3{\cal D}_{22}({\bm p})) \Big) \notag \\
    & - 2\Gamma \text{Re}\Big\{  -i\tau \chi^{-1}\psi_1   {\bm k} \cdot {\bm \pi}({\bm k}) - i \gamma \kappa^{-1}\psi_1    {\bm k} \, \hat{m}^{-1} {\bm a}({\bm k}) + 2{\bm a}({\bm k}) \hat{m}^{-1} \int\limits_{\bm p}{\bm a}({\bm p})    \Big\}.
\end{align}

{\textbf{Cross correlators}.}
\begin{align}
     \partial_t \pi_\alpha({\bm k},t) & = -\Gamma \Big[ i\tau \chi^{-1} \psi_1  \Pi_{\alpha\beta} (\bm k) k_\beta  + i\gamma\kappa^{-1}\psi_1   (K^\dagger(\bm k))_{\alpha\gamma} \hat{m}^{-1}_{\gamma\beta} k_\beta + 2 (K^\dagger(\bm k))_{\alpha\beta}\hat{m}^{-1}_{\beta\gamma} \int\limits_{\bm p} {a}_\gamma({\bm p})  \notag\\
    & + \pi_\alpha(\bm k) \Big( \alpha + \gamma^2\kappa^{-2}k_{ab}^2 + \kappa^{-2}k_z^2 + \psi_1^2 +  \int\limits_{\bm p}( \text{tr}(\hat{m}^{-1}{\Phi}({\bm p})) + {\cal D}_{11}({\bm p}) + 3{\cal D}_{22}({\bm p})) \Big)
    \Big] \notag \\
    & + \Gamma_E \Big[ (k^2\delta_{\alpha\beta} - k_\alpha k_\beta) a_\beta(\bm k) 
     - \sigma_{\alpha\beta} (\delta_{\beta\gamma} +\gamma \kappa^{-1}\chi^{-1}k_\beta k_\gamma) \pi_\gamma(\bm k)  + \hat{m}^{-1}_{\alpha\beta} a_\beta(\bm k) \Big( \psi_1^2 +  \int\limits_{\bm p}(  {\cal D}_{11}({\bm p}) 
     + {\cal D}_{22}({\bm p})) \Big) \notag\\
     & + 2{\cal D}_{22}(\bm k)\hat{m}^{-1}_{\alpha\beta}\int\limits_{\bm p} a_\beta(\bm p) + i\gamma\kappa^{-1}\psi_1\hat{m}^{-1}_{\alpha\beta} k_\beta {\cal D}_{22}(\bm k) \Big],\\
     \partial_t a_\alpha({\bm k},t) & =-\pi_\alpha(\bm k) -\Gamma \Big[ i\tau \chi^{-1} \psi_1  K_{\alpha\beta}(\bm k) k_\beta  + i\gamma\kappa^{-1}\psi_1   \Phi_{\alpha\gamma}(\bm k) \hat{m}^{-1}_{\gamma\beta} k_\beta + 2 \Phi_{\alpha\beta}(\bm k)  \hat{m}^{-1}_{\beta\gamma} \int\limits_{\bm p} {a}_\gamma({\bm p})  \notag\\
    & + a_\alpha(\bm k) \Big(  \alpha + \gamma^2\kappa^{-2}k_{ab}^2 + \kappa^{-2}k_z^2 + \psi_1^2 +  \int\limits_{\bm p}( \text{tr}(\hat{m}^{-1}{\Phi}({\bm p})) + {\cal D}_{11}({\bm p}) + 3{\cal D}_{22}({\bm p})) \Big)
    \Big].
\end{align}

{\textbf{EM sector}.}
\begin{align}
     \partial_t \Phi ({\bm k},t) & = - (K(\bm k) + K^\dagger(\bm k)), \\
     \partial_t K_{\alpha\beta} ({\bm k},t)  &= - \Pi_{\alpha\beta}(\bm k)  +\Gamma_E \Big[ \Phi_{\alpha\gamma}(\bm k) (k^2 \delta_{\gamma\beta} -k_\gamma k_\beta ) -   K_{\alpha\delta}(\bm k) ( \delta_{\delta\gamma} +\gamma \kappa^{-1}\chi^{-1} k_\delta k_\gamma)\sigma_{\gamma\beta}   \notag\\
    & + \Phi_{\alpha\gamma}(\bm k) \hat{m}^{-1}_{\gamma\beta} \Big( \psi_1^2 +  \int\limits_{\bm p}(  {\cal D}_{11}({\bm p}) + {\cal D}_{22}({\bm p})) \Big)  -i\gamma\kappa^{-1} \psi_1 a_\alpha(\bm k) \hat{m}^{-1}_{\beta \gamma} k_\gamma + 2 a_\alpha(\bm k) \hat{m}^{-1}_{\beta \gamma} \int\limits_{\bm p} a_\gamma({\bm p})\Big],\\
     \partial_t \Pi ({\bm k},t)  & = 2 Tr_0\hat{\sigma} \Gamma_E^2  + Q(\bm k) + Q^\dagger(\bm k),  \\
     Q_{\alpha\beta}(\bm k,t)  & = \Gamma_E\Big[ (K^\dagger(\bm k))_{\alpha\gamma} (k^2 \delta_{\gamma\beta} -  k_\gamma k_\beta)  -   \Pi_{\alpha\delta} (\bm k) ( \delta_{\delta\gamma} +\gamma \kappa^{-1}\chi^{-1} k_\delta k_\gamma)\sigma_{\gamma\beta}   \notag\\
    & + (K^\dagger(\bm k))_{\alpha\gamma} \hat{m}^{-1}_{ \gamma\beta}\Big( \psi_1^2 +  \int\limits_{\bm p}(  {\cal D}_{11}({\bm p}) + {\cal D}_{22}({\bm p})) \Big)  -i\gamma\kappa^{-1} \psi_1 \pi_\alpha(\bm k) \hat{m}^{-1}_{\beta \gamma} k_\gamma + 2 \pi_\alpha(\bm k) \hat{m}^{-1}_{\beta \gamma} \int\limits_{\bm p} a_\gamma({\bm p})\Big].\label{eqn:Q_full}
\end{align}

Equations~\eqref{eqn:psi_1_full}-\eqref{eqn:Q_full} represent our central technical result. In principle, using these equations, one can directly simulate a photoexcitation event. We note, however, that the total number of independent degrees of freedom is quite large, and one will have to introduce a grid in the three-dimensional momentum space, limiting simulations to relatively small system sizes. One can significantly facilitate simulations of large systems by invoking the cylindrical symmetry of anisotropic superconductors and spherical symmetry of isotropic ones. We address how to do this in practice below. We remark that the thermal state is found self-consistently by putting the right-hand sides of each of the equations to be zero.

\subsection{Cylindrical symmetry}

To take advantage of the cylindrical symmetry we use the following ansatz: 
\begin{gather}
    \bm a(\bm k,t)   = i a_{ab}(k_{ab},k_z,t)\bm k_{ab} + i a_{z}(k_{ab},k_z,t)\bm k_{z},\,
    \bm \pi(\bm k,t) = i \pi_{ab}(k_{ab},k_z,t)\bm k_{ab} + i \pi_{z}(k_{ab},k_z,t)\bm k_{z},\\
    \Phi_{\alpha\beta}(\bm k,t)  = \Phi_\parallel(k_{ab},k_z,t) \hat{T}_1 + \Phi_{X,1}(k_{ab},k_z,t) \hat{T}_2 + \Phi_{X,2}(k_{ab},k_z,t) \hat{T}_3 +\Phi_{ab}(k_{ab},k_z,t) \hat{T}_4  + \Phi_z(k_{ab},k_z,t) \hat{T}_5\label{eqn:EM_tensors},
\end{gather}
and similar expansion holds for the other two electromagnetic tensors. Here $\hat{T}_1 = k_{ab,\alpha}k_{ab,\beta}$, $\hat{T}_2 = k_{ab,\alpha}k_{z,\beta}$, $\hat{T}_3 = k_{z,\alpha}k_{ab,\beta}$, $\hat{T}_4 =\delta_{\alpha\beta}^{ab} =\delta_{\alpha,x}\delta_{\beta,x} + \delta_{\alpha,y}\delta_{\beta,y}$, and $\hat{T}_5 =\delta_{\alpha\beta}^{z} = \delta_{\alpha,z}\delta_{\beta,z}$ -- their algebra is summarized in Table~\ref{tbl:algebra}. This ansatz allows us to write a closed set of equations solely on the newly introduced quantities, thereby reducing the initial three-dimensional problem to only two-dimensional. Moreover, all of these quantities can be chosen to be real. On symmetry grounds, one has $\Phi^\dagger = \Phi$ and $\Pi^\dagger = \Pi \Rightarrow \Phi_{X,1} = \Phi_{X,2} = \Phi_{X}$ and $\Pi_{X,1} = \Pi_{X,2} = \Pi_{X}$, but in general $K_{X,1} \neq K_{X,2}$. Below we summarize the final equations of motion.

\begin{table*}[t!]
\ra{1.3}
\begin{center}
\begin{tabular}{@{}m{3cm}m{3cm}m{3cm}m{3cm}m{1.5cm}@{}}
\toprule
$\hat{T}_1\hat{T}_1 = k^2_{ab} \hat{T}_1$ & $\hat{T}_2\hat{T}_1 = 0$  & $\hat{T}_3\hat{T}_1 = k^2_{ab} \hat{T}_3$  &  $\hat{T}_4\hat{T}_1 = \hat{T}_1$& $\hat{T}_5\hat{T}_1 = 0$\\ 
$\hat{T}_1\hat{T}_2 = k^2_{ab} \hat{T}_2$ & $\hat{T}_2\hat{T}_2 = 0$  & $\hat{T}_3\hat{T}_2 = k^2_{ab}k^2_{z} \hat{T}_5$  &  $\hat{T}_4\hat{T}_2 = \hat{T}_2$ & $\hat{T}_5\hat{T}_2 = 0$\\ 
$\hat{T}_1\hat{T}_3 = 0$ & $\hat{T}_2\hat{T}_3 = k^2_{z} \hat{T}_1$  & $\hat{T}_3\hat{T}_3 = 0$  &  $\hat{T}_4\hat{T}_3 = 0$& $\hat{T}_5\hat{T}_3 = \hat{T}_3$\\ 
$\hat{T}_1\hat{T}_4 = \hat{T}_1$ & $\hat{T}_2\hat{T}_4 = 0$  & $\hat{T}_3\hat{T}_4 = \hat{T}_3$  &  $\hat{T}_4\hat{T}_4 =  \hat{T}_4$ & $\hat{T}_5\hat{T}_4 = 0$\\ 
$\hat{T}_1\hat{T}_5 = 0$ & $\hat{T}_2\hat{T}_5 =  \hat{T}_2$  & $\hat{T}_3\hat{T}_5 =0$  &  $\hat{T}_4\hat{T}_5 =0$& $\hat{T}_5\hat{T}_5 = \hat{T}_5$\\ 
\bottomrule 
\end{tabular}
\end{center}
\caption{Algebra of operators in the tensor expansion~\eqref{eqn:EM_tensors}. Note that: $\hat{T}_1^\dagger = \hat{T}_1$, $\hat{T}_2^\dagger = \hat{T}_3$, $\hat{T}_3^\dagger = \hat{T}_2$, $\hat{T}_4^\dagger = \hat{T}_4$, and $\hat{T}_5^\dagger = \hat{T}_5$.}
\label{tbl:algebra}
\end{table*}

{\textbf{SC sector}.}
\begin{align}
     \tau\partial_t \psi_1(t)  & =  \int\limits_{\bm p} (\tau \chi^{-1} (p^2_{ab} \pi_{ab}(\bm p) + p^2_z \pi_z(\bm p)) -  \gamma\kappa^{-1}(p^2_{ab} a_{ab}(\bm p) + \gamma^{-2} p^2_z a_z(\bm p)))\notag\\
     &
     - \psi_1 \Big( \alpha + \psi_1^2 +  \int\limits_{\bm p}(  p^2_{ab}\Phi_\parallel(\bm p) + 2\Phi_{ab}(\bm p) + \gamma^{-2}\Phi_z(\bm p) + 3{\cal D}_{11}({\bm p}) + {\cal D}_{22}({\bm p})) \Big),\\
     \partial_t {\cal D}_{11}({\bm k},t) & = 2 \Gamma T r_0 - 2 \Gamma {\cal D}_{11}(\bm k) \Big( \alpha + \gamma^2\kappa^{-2}k_{ab}^2 + \kappa^{-2}k_z^2 + 3 \psi_1^2\notag\\
     &
     +  \int\limits_{\bm p}(  p^2_{ab}\Phi_\parallel(\bm p) + 2\Phi_{ab}(\bm p) + \gamma^{-2}\Phi_z(\bm p) + 3{\cal D}_{11}({\bm p}) + {\cal D}_{22}({\bm p})) \Big),\\
     \partial_t {\cal D}_{22}({\bm k},t)  & = 2 \Gamma T r_0 - 2 \Gamma {\cal D}_{22}(\bm k) \Big( \alpha + \gamma^2\kappa^{-2}k_{ab}^2 + \kappa^{-2}k_z^2 + \psi_1^2 \notag\\
     & +  \int\limits_{\bm p}(  p^2_{ab}\Phi_\parallel(\bm p) + 2\Phi_{ab}(\bm p) + \gamma^{-2}\Phi_z(\bm p) + {\cal D}_{11}({\bm p}) + 3{\cal D}_{22}({\bm p})) \Big) \notag\\
     &
    - 2\Gamma\psi_1\Big[  \tau \chi^{-1} (k_{ab}^2 \pi_{ab}(\bm k) + k_z^2 \pi_z(\bm k)) + \gamma \kappa^{-1}(k_{ab}^2 a_{ab}(\bm k) + \gamma^{-2} k_z^2 a_z(\bm k))  \Big].
\end{align}

{\textbf{Cross correlators}.}
\begin{align}
     \partial_t \pi_{ab}({\bm k},t) & = - \Gamma\Big[ \tau \chi^{-1} \psi_1 (k^2_{ab}\Pi_\parallel(\bm k) + k_z^2 \Pi_{X}(\bm k)  + \Pi_{ab}(\bm k)) + \gamma\kappa^{-1}\psi_1 ( k^2_{ab}K_\parallel(\bm k) + \gamma^{-2}k^2_z K_{X,2}(\bm k) + K_{ab}(\bm k))\notag
    \\
    & + \pi_{ab}(\bm k) \Big(  \alpha + \gamma^2\kappa^{-2}k_{ab}^2 + \kappa^{-2}k_z^2 + \psi_1^2 +  \int\limits_{\bm p}( p^2_{ab}\Phi_\parallel(\bm p) + 2\Phi_{ab}(\bm p) + \gamma^{-2}\Phi_z(\bm p) + {\cal D}_{11}({\bm p}) + 3{\cal D}_{22}({\bm p})) \Big)\Big] \notag\\
     &
    + \Gamma_E\Big[ k^2_z (a_{ab}(\bm k) - a_z(\bm k))
    - \sigma_{ab}(\pi_{ab} (\bm k) + \gamma\kappa^{-1}\chi^{-1}(k_{ab}^2 \pi_{ab}(\bm k) + k_{z}^2 \pi_{z}(\bm k))) \notag\\
    &
    + a_{ab}(\bm k) \Big( \psi_1^2 +  \int\limits_{\bm p}(  {\cal D}_{11}({\bm p}) 
     + {\cal D}_{22}({\bm p})) \Big) + \gamma\kappa^{-1}\psi_1 {\cal D}_{22}(\bm k) \Big],
\end{align}
\begin{align}
    \partial_t \pi_{z}({\bm k},t) & =  - \Gamma\Big[ \tau \chi^{-1} \psi_1 (k^2_{ab} \Pi_{X}(\bm k) + \Pi_z(\bm k)) + \gamma\kappa^{-1}\psi_1 (k^2_{ab} K_{X,1}(\bm k) + \gamma^{-2}K_z(\bm k)) \notag
    \\
    & + \pi_z(\bm k) \Big(  \alpha + \gamma^2\kappa^{-2}k_{ab}^2 + \kappa^{-2}k_z^2 + \psi_1^2 +  \int\limits_{\bm p}( p^2_{ab}\Phi_\parallel(\bm p) + 2\Phi_{ab}(\bm p) + \gamma^{-2}\Phi_z(\bm p) + {\cal D}_{11}({\bm p}) + 3{\cal D}_{22}({\bm p})) \Big)\Big] \notag\\
    & + \Gamma_E\Big[ k^2_{ab} (a_{z}(\bm k) - a_{ab}(\bm k)) - \sigma_{c}(\pi_{z}(\bm k) + \gamma\kappa^{-1}\chi^{-1}(k_{ab}^2 \pi_{ab}(\bm k) + k_{z}^2 \pi_{z}(\bm k))) \notag\\
    &+ \gamma^{-2} a_{z}(\bm k) \Big( \psi_1^2 +  \int\limits_{\bm p}(  {\cal D}_{11}({\bm p}) 
     + {\cal D}_{22}({\bm p})) \Big) + \gamma^{-1}\kappa^{-1}\psi_1 {\cal D}_{22}(\bm k)\Big],
\end{align}
\begin{align}   
    \partial_t a_{ab}({\bm k},t) & = - \pi_{ab}(\bm k) - \Gamma\Big[ \tau \chi^{-1} \psi_1 (k^2_{ab}K_\parallel(\bm k) + k_z^2 K_{X,1}(\bm k)  + K_{ab}(\bm k)) + \gamma\kappa^{-1}\psi_1 ( k^2_{ab}\Phi_\parallel(\bm k) + \gamma^{-2}k^2_z\Phi_{X}(\bm k) + \Phi_{ab}(\bm k))\notag
    \\
    & + a_{ab}(\bm k) \Big(  \alpha + \gamma^2\kappa^{-2}k_{ab}^2 + \kappa^{-2}k_z^2 + \psi_1^2 +  \int\limits_{\bm p}( p^2_{ab}\Phi_\parallel (\bm p) + 2\Phi_{ab}(\bm p) + \gamma^{-2}\Phi_z(\bm p) + {\cal D}_{11}({\bm p}) + 3{\cal D}_{22}({\bm p})) \Big)\Big],\\
    \partial_t a_{z}({\bm k},t) & = - \pi_z(\bm k) - \Gamma\Big[ \tau \chi^{-1} \psi_1 (k^2_{ab} K_{X,2}(\bm k) + K_z(\bm k)) + \gamma\kappa^{-1}\psi_1 (k^2_{ab} \Phi_{X}(\bm k) + \gamma^{-2}\Phi_z(\bm k)) \notag
    \\
    & + a_z(\bm k)\Big(  \alpha + \gamma^2\kappa^{-2}k_{ab}^2 + \kappa^{-2}k_z^2 + \psi_1^2 +  \int\limits_{\bm p}( p^2_{ab}\Phi_\parallel(\bm p) + 2\Phi_{ab}(\bm p) + \gamma^{-2}\Phi_z(\bm p) + {\cal D}_{11}({\bm p}) + 3{\cal D}_{22}({\bm p})) \Big)\Big].
\end{align}

{\textbf{EM sector}.}
\begin{align}
    \partial_t \Phi_\parallel(\bm k, t) & = -2 K_\parallel(\bm k),\, \partial_t \Phi_{X}(\bm k, t)  = -( K_{X,1}(\bm k) + K_{X,2}(\bm k)),\, \partial_t \Phi_{ab}(\bm k, t)  = -2 K_{ab}(\bm k),\, \partial_t \Phi_{z}(\bm k, t)  = -2 K_{z}(\bm k),\\
    \partial_t K_\parallel(\bm k, t) & = -\Pi_\parallel(\bm k) + \Gamma_E\Big[ \Big(  k_z^2\Phi_\parallel(\bm k)  - k_{z}^2\Phi_{X}(\bm k) - \Phi_{ab}(\bm k) \Big) +  \Big( \psi_1^2 +  \int\limits_{\bm p}(  {\cal D}_{11}({\bm p}) + {\cal D}_{22}({\bm p})) \Big) \Phi_\parallel(\bm k) \notag\\
    &- \sigma_{ab}\Big(  K_\parallel(\bm k) + \frac{\gamma}{\kappa\chi}(k^2_{ab} K_\parallel(\bm k) + k^2_{z} K_{X,1}(\bm k) + K_{ab}(\bm k) ) \Big)
     + \gamma\kappa^{-1} \psi_1 a_{ab}(\bm k)\Big],
    \\
    \partial_t K_{X,1}(\bm k, t) & = -\Pi_{X}(\bm k) + \Gamma_E\Big[ \Big( k_{ab}^2\Phi_{X}(\bm k) - k_{ab}^2\Phi_\parallel(\bm k)  - \Phi_{ab}(\bm k) \Big) + \gamma^{-2} \Big( \psi_1^2 +  \int\limits_{\bm p}(  {\cal D}_{11}({\bm p}) + {\cal D}_{22}({\bm p})) \Big) \Phi_{X}(\bm k)\notag\\
    &
    - \sigma_{c}\Big(  K_{X,1}(\bm k) + \frac{\gamma}{\kappa\chi}( k^2_{ab} K_\parallel(\bm k) + k_z^2 K_{X,1}(\bm k) + K_{ab}(\bm k) )  \Big) 
     + \gamma^{-1}\kappa^{-1} \psi_1 a_{ab}(\bm k)\Big],
    \\
    \partial_t K_{X,2} (\bm k, t) & = - \Pi_{X}(\bm k) + \Gamma_E\Big[ \Big(  k_z^2\Phi_{X}(\bm k) - \Phi_z(\bm k)\Big) + \Big( \psi_1^2 +  \int\limits_{\bm p}(  {\cal D}_{11}({\bm p}) + {\cal D}_{22}({\bm p})) \Big) \Phi_{X}(\bm k) \notag\\
    &
     - \sigma_{ab}\Big(  K_{X,2}(\bm k) + \frac{\gamma}{\kappa\chi}( k^2_{ab} K_{X,2}(\bm k) + K_z(\bm k))  \Big) + \gamma\kappa^{-1} \psi_1 a_{z}(\bm k)\Big],
    \\
    \partial_t K_{ab}(\bm k, t) & = - \Pi_{ab}(\bm k) + \Gamma_E\Big[  (k_{ab}^2 + k_z^2)\Phi_{ab}(\bm k)  - \sigma_{ab} K_{ab}(\bm k)  +  \Big( \psi_1^2 +  \int\limits_{\bm p}(  {\cal D}_{11}({\bm p}) + {\cal D}_{22}({\bm p})) \Big) \Phi_{ab}(\bm k)\Big],
    \\
    \partial_t K_{z}(\bm k, t) & =- \Pi_{z}(\bm k) + \Gamma_E\Big[ \Big( k_{ab}^2 \Phi_z(\bm k) - k_{ab}^2k_z^2 \Phi_{X}(\bm k)\Big) +  \gamma^{-2}\Big( \psi_1^2 +  \int\limits_{\bm p}(  {\cal D}_{11}({\bm p}) + {\cal D}_{22}({\bm p})) \Big) \Phi_{z}(\bm k) \notag\\
    &
    -  \sigma_{c}\Big( K_{z}(\bm k) + \frac{\gamma}{\kappa\chi}( k^2_{ab} k^2_z K_{X,2}(\bm k) + k_z^2 K_z(\bm k) )  \Big) + \gamma^{-1}\kappa^{-1} \psi_1 a_{z}(\bm k) k_z^2\Big],
\end{align}
\begin{align}
    \partial_t \Pi_\parallel(\bm k, t) & = 2\Gamma_E\Big[ \Big(  k_z^2K_\parallel(\bm k)  - k_{z}^2 K_{X,2}(\bm k) - K_{ab}(\bm k) \Big) +  \Big( \psi_1^2 +  \int\limits_{\bm p}(  {\cal D}_{11}({\bm p}) + {\cal D}_{22}({\bm p})) \Big) K_\parallel(\bm k) \notag\\
    &
     - \sigma_{ab}\Big(  \Pi_\parallel(\bm k) + \frac{\gamma}{\kappa\chi}(k^2_{ab} \Pi_\parallel(\bm k) + k^2_{z} \Pi_{X}(\bm k) + \Pi_{ab}(\bm k) ) \Big) + \gamma\kappa^{-1} \psi_1 \pi_{ab}(\bm k)\Big],
\end{align}
\begin{align}
    \partial_t \Pi_{X}(\bm k, t) & = \Gamma_E\Big[ \Big( (k_{ab}^2 + k_z^2)(K_{X,1}(\bm k) + K_{X,2}(\bm k)) - k_{ab}^2 K_\parallel(\bm k) - k_{z}^2 K_{X,2}(\bm k) - K_{ab}(\bm k) - K_z(\bm k) - k_{ab}^2 K_{X,1}(\bm k) \Big) \notag\\
    &
    - \sigma_{ab}\Big(  \Pi_{X}(\bm k) + \frac{\gamma}{\kappa\chi}( k^2_{ab} \Pi_{X}(\bm k) + \Pi_z (\bm k))  \Big) - \sigma_{c}\Big(  \Pi_{X}(\bm k) + \frac{\gamma}{\kappa\chi}( k^2_{ab} \Pi_\parallel(\bm k) + k_z^2 \Pi_{X}(\bm k) + \Pi_{ab}(\bm k) )  \Big) \notag\\
    & +  \gamma\kappa^{-1} \psi_1 (\pi_z(\bm k) + \gamma^{-2}\pi_{ab}(\bm k))
      + \Big( \psi_1^2 +  \int\limits_{\bm p}(  {\cal D}_{11}({\bm p}) + {\cal D}_{22}({\bm p})) \Big) (K_{X,1}(\bm k) + \gamma^{-2}K_{X,2}(\bm k)) \Big],\\
    %
    %
% \end{align}
% \begin{align}
    \partial_t \Pi_{ab}(\bm k, t) & = 2Tr_0\sigma_{ab} \Gamma^2_E + 2\Gamma_E\Big[   (k_{ab}^2 + k_z^2)K_{ab}(\bm k)  - \sigma_{ab} \Pi_{ab}(\bm k)  +  \Big( \psi_1^2 +  \int\limits_{\bm p}(  {\cal D}_{11}({\bm p}) + {\cal D}_{22}({\bm p})) \Big) K_{ab}(\bm k)\Big],
    \\
    \partial_t \Pi_{z}(\bm k, t) & =2Tr_0 \sigma_c \Gamma^2_E + 2\Gamma_E\Big[ \Big( k_{ab}^2 K_z(\bm k) - k_{ab}^2k_z^2 K_{X,1}(\bm k) \Big) -  \sigma_{c}\Big( \Pi_{z}(\bm k)+ \frac{\gamma}{\kappa\chi}( k^2_{ab} k^2_z \Pi_{X}(\bm k) + k_z^2 \Pi_z(\bm k) )  \Big)\notag\\
    &
    +  \gamma^{-2}\Big( \psi_1^2 +  \int\limits_{\bm p}(  {\cal D}_{11}({\bm p}) + {\cal D}_{22}({\bm p})) \Big) K_{z}(\bm k) + \gamma^{-1}\kappa^{-1} \psi_1 \pi_{z}(\bm k) k_z^2\Big].
\end{align}

\subsection{Spherical symmetry}

For isotropic three-dimensional superconductors, with $\sigma_{ab} = \sigma_c = \sigma$ and $\gamma = 1$, one simplifies the equations of motion using the following ansatz:
\begin{align}
    \bm a(\bm k,t) & = i a_k(t) \bm k,\, \bm \pi(\bm k) = i \pi_k(t) \bm k,\\
    \Phi_{\alpha\beta}(\bm k,t) & = \Phi^\parallel_k(t) \frac{k_\alpha k_\beta}{k^2} + \Phi^\perp_k(t)\Big(\delta_{\alpha\beta} - \frac{k_\alpha k_\beta}{k^2}\Big),
\end{align}
and similar expansion holds for the other two electromagnetic tensors. With this ansatz, the initial three-dimensional problem reduces to one-dimensional. All of the newly introduced quantities are real.

The final set of equations of motion for isotropic superconductors reads:
\begin{align}
    \tau \partial_t \psi_1  & = \int\limits_{\bm p}  p^2 \left[ \tau \chi^{-1} \pi_p - \kappa^{-1}  a_p \right] - \psi_1 \Big[ \alpha + \psi_1^2 + \int\limits_{\bm p}  (\Phi^\parallel_p +2\Phi^\perp_p + 3 {\cal D}_{11}( p) + {\cal D}_{22}( p)) \Big],\\
     \tau \partial_t {\cal D}_{11}(k,t)  & = 2 T r_0 - 2  {\cal D}_{11}(k) \Big[ \alpha + \kappa^{-2}k^2 + 3 \psi_1^2 + \int\limits_{\bm p}  (\Phi^\parallel_p +2\Phi^\perp_p + 3 {\cal D}_{11}(p) + {\cal D}_{22}(p) )\Big],\\
    \tau  \partial_t {\cal D}_{22}(k,t)  & = 2 T r_0 - 2  {\cal D}_{22} \Big[\alpha + \kappa^{-2}k^2 + \psi_1^2 + \int\limits_{\bm p}  (\Phi^\parallel_p +2\Phi^\perp_p + {\cal D}_{11}(p) + 3{\cal D}_{22}(p))\Big] - 2 \psi_1 k^2 [  \tau  \chi^{-1}\pi_k   + \kappa^{-1} a_k  ],\\
      \partial_t \Phi^{\parallel(\perp)}_k(t) & = -2 K^{\parallel(\perp)}_k,\\
     \partial_t K^\parallel_k(t) &  = - \Pi^\parallel_k  +\Gamma_E \Big[ - \sigma   K^\parallel_k ( 1 +\kappa^{-1}\chi^{-1} k^2) + \Phi^\parallel_k \Big( \psi_1^2 +  \int\limits_{\bm p}({\cal D}_{11}({ p}) + {\cal D}_{22}({ p})) \Big)  + \kappa^{-1} \psi_1 k^2 a_k  \Big],\\
     \partial_t K^\perp_k(t) & = - \Pi^\perp_k  +\Gamma_E \Big[ - \sigma   K^\perp_k + \Phi^\perp_k \Big( k^2 + \psi_1^2 +  \int\limits_{\bm p}({\cal D}_{11}({ p}) + {\cal D}_{22}({ p})) \Big)  \Big],
\end{align}
\begin{align}
     \partial_t \Pi^\parallel_k(t) &  = 2 Tr_0\sigma \Gamma_E^2  + 2\Gamma_E \Big[ - \sigma   \Pi^\parallel_k ( 1 +\kappa^{-1}\chi^{-1} k^2) + K^\parallel_k \Big( \psi_1^2 +  \int\limits_{\bm p}({\cal D}_{11}({ p}) + {\cal D}_{22}({ p})) \Big)  + \kappa^{-1} \psi_1 k^2 \pi_k  \Big],\\
     \partial_t \Pi^\perp_k(t) & = 2 Tr_0\sigma \Gamma_E^2 +2 \Gamma_E \Big[ - \sigma   \Pi^\perp_k + K^\perp_k \Big( k^2 + \psi_1^2 +  \int\limits_{\bm p}({\cal D}_{11}({ p}) + {\cal D}_{22}({ p})) \Big)  \Big],
\end{align}
\begin{align}
     \partial_t \pi_k(t) & = -\Gamma \Big[ 
     \tau \chi^{-1} \psi_1  \Pi_k^\parallel 
    +\kappa^{-1} \psi_1 K_k^\parallel + \pi_k \Big( \alpha + \kappa^{-2}k^2 + \psi_1^2 +  \int\limits_{\bm p}  (\Phi^\parallel_p +2\Phi^\perp_p + {\cal D}_{11}(p) + 3 {\cal D}_{22}(p)) \Big) \Big]\notag\\
    &\qquad\qquad\qquad +\Gamma_E\Big[ a_k\Big( \psi_1^2 +  \int\limits_{\bm p}  ({\cal D}_{11}(p) + {\cal D}_{22}(p)) \Big) - \sigma( 1 + \kappa^{-1}\chi^{-1}k^2)\pi_k + \kappa^{-1}  \psi_1 {\cal D}_{22}(k)\Big],\\
     \partial_t a_k(t) & = -  \pi_k - \Gamma \Big[ 
     \tau \chi^{-1} \psi_1  K_k^\parallel 
    +\kappa^{-1} \psi_1 \Phi_k^\parallel + a_k \Big( \alpha + \kappa^{-2}k^2 + \psi_1^2 +  \int\limits_{\bm p}  (\Phi^\parallel_p +2\Phi^\perp_p + {\cal D}_{11}(p) + 3 {\cal D}_{22}(p)) \Big) \Big].
\end{align}

\section{Dynamics of the order parameter fluctuations after photoexcitation in the symmetry broken phase}
\label{appendix:broken_phase}

\begin{figure}[t!]
	\centering
	\includegraphics[width=0.8\linewidth]{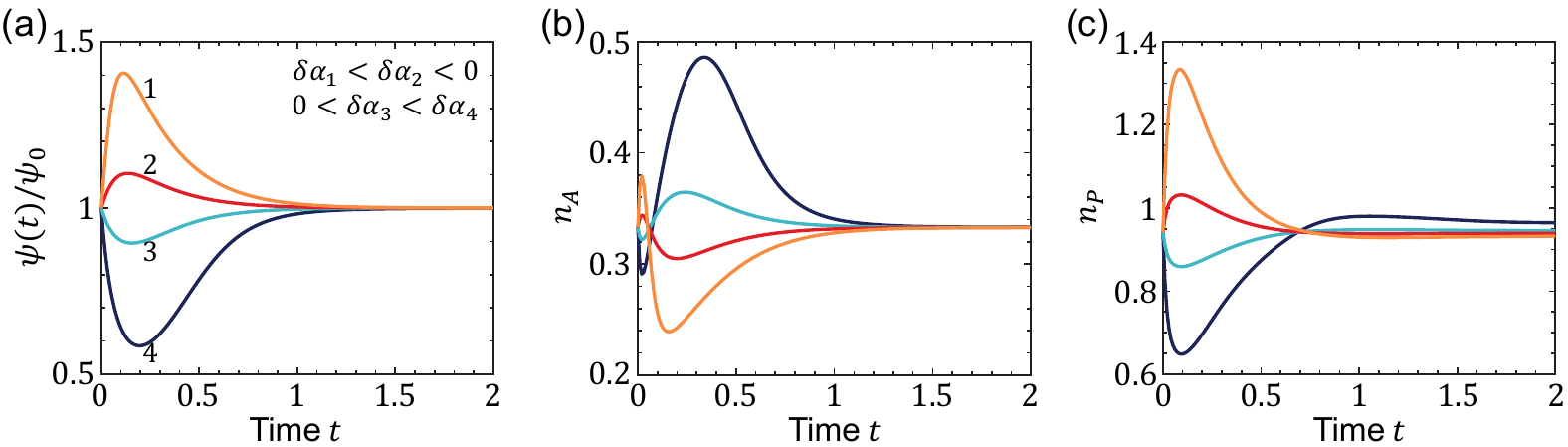}
	\caption{Photoexcitation dynamics in isotropic superconductors below $T_c$ -- extension of Fig.~\ref{fig:Quenches} (bottom panels) of the main text. (a) Dynamics of the long-range order parameter expectation value $\langle \psi \rangle(t)$: for $\delta \alpha < 0$ ($\delta \alpha > 0$), it becomes transiently enhanced (suppressed) and then exponentially returns to its pre-pulse value $\psi_0$. We decompose $\langle |\psi|^2 \rangle(t) = \psi^2(t) + n_A(t) + n_P(t)$, where $n_A(t)$ represents the longitudinal order parameter fluctuations (b) and $n_P(t)$ describes the transverse fluctuations (c).
	}
\label{fig:S1}
\end{figure}

When considering quenches in the symmetry broken phase in the main text, we showed only the dynamics of $\langle |\psi|^2\rangle(t)$, which actually contains contributions from both the long-range expectation value $\langle\psi\rangle$ and order parameter fluctuations: $\langle |\psi|^2 \rangle = \langle \psi \rangle^2 + n_A + n_P$. Here $n_A(t) = \int_{\bm p} {\cal D}_{11}(p,t)$ represents longitudinal order parameter fluctuations and encompasses the order parameter amplitude; $n_P(t) = \int_{\bm p} {\cal D}_{22}(p,t)$ describes transverse fluctuations and encodes essentially the order parameter phase. Figure~\ref{fig:S1} shows the dynamics of each of these quantities. 

For concreteness, we stick to quenches with $\delta \alpha < 0$ corresponding to photo-enhancement of superconductivity (yellow and red curves in Fig.~\ref{fig:S1}). The order parameter dynamics is similar to that of $\langle |\psi|^2 \rangle(t)$: $\langle \psi \rangle$ is first transiently enhanced and then exponentially restores to its pre-pulse value $\psi_0$. The stronger the photoexcitation, the stronger the order parameter develops.

The evolution of the longitudinal fluctuations is governed by three stages. During the first quick stage, $n_A(t)$ slightly proliferates because transiently the Ginzburg-Landau free energy becomes steeper. During the second stage, $n_A(t)$ becomes suppressed due to the development of the order parameter expectation value $\langle \psi \rangle (t)$, which renders amplitude fluctuations energetically costly. The final stage is the recovery to the equilibrium state. 

The evolution of the transverse fluctuations is different: initially, when the free energy becomes steeper, $n_P(t)$ proliferates and then seemingly recovers to its equilibrium value, following the trend of $\langle \psi\rangle(t)$. However, before actually recovering, $n_P(t)$ transiently becomes slightly suppressed compared to its equilibrium value. This final stage can be understood as follows.
In contrast to the amplitude fluctuations, the phase fluctuations are linearly coupled to the electromagnetic field, resulting in the development of a plasma gap. Thus, the initial rise of $n_P(t)$ can be interpreted as a proliferation of plasmons at all length scales. Since at longer times $\langle |\psi|^2 \rangle(t)$ exceeds its equilibrium value and since this quantity defines the plasmon frequency at equilibrium, it renders the plasmons to be energetically costly, resulting in their eventual depopulation.

\section{Dephasing within the Scenario I}
\label{appendix:Scenario I}

In the main text, we primarily studied the situation where the photoexcitation results in a sudden quench of the quadratic coefficient $\alpha(t)$ of the superconducting free energy. Although $\alpha(t)$ shows abrupt dynamics, the evolution of the superconducting order parameter is relatively smooth. As it evolves, it excites the entire plasmon continuum through the generation of momentum conserving plasmon pairs. For smoother order parameter dynamics, the high energy, high momentum modes are less excited. For this reason, the dephasing effect is relatively weak and many cycles of bi-plasmon oscillations are visible before their decay. 

In the phenomenological model, if one chooses an extremely large order parameter relaxation time, as encoded in $\tau$, then the resulting bi-plasmon oscillations will be suppressed. This is because even the low momenta plasmons are not being excited in this regime, which is the case of adiabatic order parameter dynamics. On the contrary, if $\tau$ is extremely small, then the order parameter displays abrupt evolution. In this case, the amplitude of bi-plasmon oscillations is large, but their lifetime is small due to dephasing. This damping is also stronger for superconductors with small Ginzburg parameter $\kappa$ and, as such, steeper plasmon dispersion, cf. Appendix~\ref{sec: Collective modes}.

\begin{figure}[tbh!]
	\centering
	\includegraphics[width=0.5\linewidth]{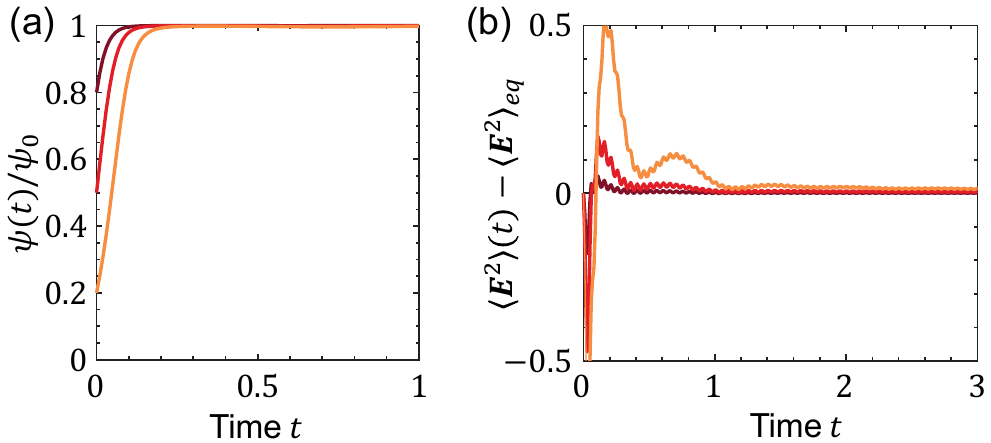}
	\caption{ Post-pulse dynamics within the Scenario I in isotropic superconductors below $T_c$. To mimic the photoexcitation event, we choose the initial state to be thermal, but then we take the order parameter expectation value $\psi(t = 0^+)$ to be reduced compared to the pre-pulse value $\psi_0$. (a) Order parameter $\psi(t)$ displays exponential recovery to the equilibrium value $\psi_0$. (b) The electric field variance, $\langle \bm E^2 \rangle(t)-\langle \bm E^2 \rangle_{eq}$, shows periodic dynamics with frequency being twice the plasmon gap. Notably, the lifetime of the oscillations here is smaller compared to the more smooth quenches considered in the main text.
	Parameters used: $\tau = 1$, $\tau_E = 1$, $\chi^{-1} = 0.1$, $\kappa = 5$, $\sigma = 0.1$, $Tr_0 = 10^{-2}$.
	%$\alpha_0 = -25$.
	}
\label{fig:S2}
\end{figure}

To illustrate the dephasing effect, we consider a situation where the photoexcitation partially  evaporates the equilibrium order parameter in a sudden manner, and we choose relatively small $\kappa$. Specifically, we prepare the initial state to be thermal below $T_c$, but then we choose $\langle \psi\rangle (t = 0^+)$ to be smaller than the equilibrium value $\psi_0$. As such, the order parameter dynamics is abrupt [Fig.~\ref{fig:S2}(a)], resulting in a rather strong damping of the bi-plasmon oscillations [Fig.~\ref{fig:S2}(b)]. We remark that the evolution in Fig.~\ref{fig:S2}(b) also displays superficial high-frequency oscillations. Those oscillations arise due to the momentum cutoff chosen in our simulation and indicate that all plasmons up to the highest momentum modes are notably excited by the abrupt change in the order parameter.

\section{Induced Periodic dynamics in regime III}
\label{appendix:regime_3}

For temperatures $T > T^*$, plasmons are overdamped. However, one can imagine that an impulsive optical quench can induce periodic dynamics. If the order parameter relaxation rate is low, then photoexcitation can result in a transient enhancement of superconducting fluctuations, which only slowly recover to equilibrium.  The developed expectation value $\langle |\psi|^2\rangle(t)$ provides the necessary ground to form lasting bi-plasmon oscillations in out-of-equilibrium. As shown in Fig.~\ref{fig:Quenches}(b) of the main text, we indeed find that such a quench results in oscillatory dynamics of the electromagnetic field. Fourier analysis of those oscillations indicates that the frequency is rather poorly defined. This is because the quasiparticle conductivity is large, so that those out-of-equilibrium oscillations are damped, and also because $\langle |\psi|^2\rangle(t)$ is a time-evolving quantity, which translates as the plasmon frequency is being changed in time.

\end{document}